\documentstyle[11pt,aaspp4]{article}
\font\qub=cmr10 scaled 1000
\begin{document}
\title{SBS 0335--052, A PROBABLE NEARBY YOUNG DWARF GALAXY: EVIDENCE PRO AND CON
\footnote{Spectroscopic
observations presented herein were obtained with the Multiple Mirror
Telescope, a facility operated jointly by the Smithsonian Institution
and the University of Arizona.}}

\author{Yuri I. Izotov}
\affil{Main Astronomical Observatory, Ukrainian National Academy of Sciences,
Goloseevo, Kiev 252650, Ukraine \\ Electronic mail: izotov@mao.gluk.apc.org}
\and
\author{Valentin A. Lipovetsky}
\affil{Special Astrophysical Observatory, Russian Academy of Sciences,
Nizhny Arkhyz, Karachai-Circessia 357147, Russia \\ Electronic mail: val@sao.ru}
\and
\author{Frederic H. Chaffee and Craig B. Foltz}
\affil{Multiple Mirror Telescope Observatory, University of Arizona, 
Tucson, AZ 85721 \\ Electronic mail: fchaffee@keck.hawaii.edu, cfoltz@as.arizona.edu}
\and
\author{Natalia G. Guseva}
\affil{Main Astronomical Observatory, Ukrainian National Academy of Sciences,
Goloseevo, Kiev 252650, Ukraine \\ Electronic mail: guseva@mao.gluk.apc.org}
\and
\author{Alexei Y. Kniazev}
\affil{Special Astrophysical Observatory, Russian Academy of Sciences,
Nizhny Arkhyz, Karachai-Circessia 357147, Russia \\ Electronic mail: akn@sao.ru}

\newpage

\begin{abstract}
  
The results of Multiple Mirror Telescope (MMT) spectrophotometry 
of the extremely low-metallicity blue compact
galaxy (BCG) SBS 0335--052 (SBS -- the Second Byurakan Survey) are presented. 
The oxygen
abundance in central brightest part of the galaxy is found to be 12 + log(O/H) =
7.33$\pm$0.01, only slightly greater than the oxygen abundance in the 
most metal-deficient
BCG I Zw 18. We show that the N/O, Ne/O, S/O and Ar/O abundance ratios in
SBS 0335--052 are close to those derived in other BCGs, suggesting that heavy
element enrichment in the HII region is due to massive star evolution. 
However, we find an O/Fe abundance ratio close to that in the Sun, in variance
with values derived for other BCGs.
The helium abundance derived from the HeI $\lambda$4471, 5876 and 6678 emission 
lines, taking into account of collisional and fluorescent enhancement, is
Y = 0.245$\pm$0.006, close to the value of the primordial helium
abundance Y$_p$=0.243$\pm$0.003 derived by Izotov, Thuan \& Lipovetsky.
We detect auroral [OIII]$\lambda$4363 emission in the inner part of HII
region with a diameter of 14$''$ or 3.6 kpc and find that the HII 
region inside this diameter is hot, $T_e$$\sim$20000K. The oxygen abundance
in this region is nearly constant (12 + log(O/H) = 7.1 -- 7.3) with a gradual
decrease to the outer part of HII region, implying effective mixing of ionized
gas on short time-scales.
We study the distribution of the nebular HeII $\lambda$4686 
emission line and find it is not produced by main-sequence O-stars or Wolf-Rayet
stars. Possible excitation mechanisms of this line,
such as massive X-ray binaries and shocks, are discussed.
We also discuss the origin of blue underlying extended low-intensity emission
detected in SBS 0335--052 on $V$, $R$ and $I$ images. 
The blue $(V-I)$ and $(R-I)$ color distributions suggest that
a significant contribution to the extended low-intensity envelope is
due to ionized gas emission. This is evidence that SBS 0335--052 is a
young galaxy experiencing its very first burst of star formation. However, we
find that the observed equivalent width of H$\beta$ emission
in the extended envelope is 2--3 times lower than the value expected in 
the case of
pure gaseous emission. Furthermore, we find that the widths of H$\gamma$ and
H$\beta$ are narrower than the instrumental profiles; this could be
explained by presence of underlying stellar absorption from A stars. These
findings suggest that, along with the blue young ($\sim$10$^7$yr) stellar 
clusters in the center of the galaxy,
an older stellar population with age $\sim$10$^8$yr may be
present in the extended envelope of
SBS 0335--052, having a total mass $\sim$ 
10$^7$M$_\odot$, two orders of magnitude smaller than the neutral gas mass
but comparable with the total mass of stars in blue young stellar clusters
observed in the center of the galaxy. We conclude that SBS 0335--052 is a
young nearby dwarf galaxy with age $\sim$ 10$^8$yr.

\end{abstract}


\section{INTRODUCTION}

The blue compact galaxy (BCG) SBS 0335--052 (SBS -- the Second Byurakan Survey)
was shown by Izotov et al. (1990ab) to have extremely low
oxygen abundance: 12+log (O/H) = 7.0 -- 7.1.  This galaxy has 
equatorial coordinates $\alpha$(1950) =
03$^h$35$^m$15.2$^s$, $\delta$(1950) = --05$^\circ$12$'$26$''$, apparent
magnitude $V$ = 16.65 mag (Thuan, Izotov \& Lipovetsky 1996), redshift 
$z$ = 0.0136 and absolute magnitude $M_V$=--17.02. Subsequent spectrophotometric
observations of SBS 0335--052 confirmed the very low metallicity in its HII
region. Terlevich et al. (1992) have derived 12+log (O/H) = 7.26, similar to
that of I Zw 18, the lowest metal-deficient BCG known. Melnick, 
Heydari-Malayeri \& Leisy (1992) have obtained $B$, $V$, $R$ 
and H$\alpha$ images of
SBS 0335--052 and found that the galaxy consists of 2 knots,
labeled A and B, separated
by 1.0$''$. The diameter of the HII region derived from the H$\alpha$
image is $\sim$ 3 kpc and shows wisps and filaments. These
authors have derived an oxygen abundance 12 + log (O/H) =
7.64 and 7.30 for knots A and B. The oxygen abundance in knot B is close
to the value derived by Skillman \& Kennicutt (1993) for I Zw 18. 

The low heavy-element abundance in SBS 0335--052 implies that this 
galaxy could be a nearby
young galaxy experiencing its first burst of star formation. Several new studies
found additional evidence for the youth of SBS 0335--052. Garnett et al.
(1995) have derived the carbon abundance in this galaxy from Hubble Space 
Telescope (HST) $UV$ spectra and
found a C/O abundance ratio $\sim$ 5 times lower than that in the Sun. Such 
a low
C/O ratio is expected from models of chemical evolution for low-metallicity, 
unevolved galaxies (Carigi et al. 1994). 
From HST $V$ and $I$ images Thuan, Izotov \& Lipovetsky (1996)
have found that SBS 0335--052 consists of 6 compact 
luminous blue clusters surrounded by extended ($\sim$ 3 kpc) low-intensity
emission elongated in the SE--NW direction with $(V-I)$ color not 
redder than 0.1--0.2, the color of late B or early A stars. 
However, its filamentary
structure may preclude a stellar origin for this emission. 
From Very Large Array (VLA) observations in the HI 21 cm emission line
(Thuan et al. 1996), a large 64kpc$\times$34kpc 
neutral gas cloud was found in the SBS 0335--052 
region elongated in the direction E--W. 
Neutral gas mass estimates give M$_{HI}$$\geq$10$^9$M$_\odot$,
which is significantly larger than the mass of the stellar population or of the
ionized gas ($\sim$10$^7$--10$^8$M$_\odot$). The VLA map shows two peaks in the 
HI distribution separated by 24 kpc. The optical counterpart of SBS 0335--052 
coincides with the eastern peak, while the western peak is found to coincide 
with a faint ($\sim$ 2 mag fainter) HII region which has been shown 
(Pustilnik et al. 1996;
Lipovetsky et al. 1996) to have the same redshift as SBS 0335--052 and a very low
oxygen abundance. Therefore, observational evidence suggests that
SBS 0335--052 may be a young nearby galaxy, a distinction it shares with
I Zw 18. However, even though it has nearly the same oxygen abundance, 
SBS 0335--052 is $\sim$ 3 mag more luminous than I Zw 18 
and contains a larger HII region and a larger number of young massive stars.

Although it is firmly established from optical spectrophotometric 
observations that SBS 0335--052 is an extremely metal-deficient galaxy,
detailed spectrophotometric studies in the optical range have yet to be
undertaken. Such observations are important in that they allow us to 
estimate the primordial helium abundance of extremely low-metallicity BCGs. 
Izotov, Thuan \& Lipovetsky (1996) have shown that the primordial helium 
abundance cannot be estimated accurately for I Zw 18 because of the 
peculiarity
of its HeI emission line intensities. Hence, SBS 0335--052 is currently the 
most 
metal-deficient BCG for which the helium abundance can be measured reliably.
The helium mass fraction, Y, in SBS 0335--052 given by Terlevich et al.
(1992) is based on measurements of the HeI $\lambda$4471 emission 
line and has a very
low value Y=0.215, while Melnick, Heydari-Malayeri \& Leisy (1992) have
derived Y=0.211 and 0.233 for knots A and B, respectively. These values of the
helium abundance are lower than that derived by Izotov, Thuan \&
Lipovetsky (1994, 1996) for other low-metallicity BCGs. 

In addition, spectrophotometric observations of SBS 0335--052 and 
the determination of the abundances of heavy elements
are important to the study of the origin of heavy elements in a
low-metallicity environment. Of further interest is the question of the
origin of the low-intensity extended emission in SBS 0335--052 which was 
detected by Thuan,
Izotov \& Lipovetsky (1996) in HST $V$ and $I$ images and of whether, as 
they suggested, dust exists in SBS 0335--052. To address
these questions we have obtained new high signal-to-noise (S/N) 
spectrophotometric observations of SBS 0335--052 and use $V$, $R$, $I$ 
photometry obtained by Thuan, Izotov \& Lipovetsky (1996) and Lipovetsky et al.
(1996).

In \S2 we present a description of the observations and data reduction; in 
\S3 the physical conditions and chemical composition in HII region are discussed. 
In \S4 we discuss possible mechanisms which could lead to the 
appearence of the strong
nebular HeII $\lambda$4686 emission line; in \S5 the kinematic 
properties of HII region are discussed; and in \S6 we model the color 
distribution of the underlying low-intensity extended component. 
We summarize our results in \S7.

\section{OBSERVATIONS AND DATA REDUCTION}

Spectrophotometric observations of SBS 0335--052 with signal-to-noise
ratios of S/N = 30 in the continuum were obtained with the Multiple
Mirror Telescope (MMT) on the night of 1995, December 20. 
Observations were made with the Blue Channel of the MMT Spectrograph
using a highly-optimized Loral 3072$\times$1024 CCD detector. A
1$''$$\times$ 180$''$ slit was used along with a 500 g/mm grating in
first order and an L--38 second order blocking filter. This gives a spatial
scale along the slit of 0.3 arcsec\ pixel$^{-1}$, a scale perpendicular
to the slit of 1.9 \AA\ pixel$^{-1}$, a spectral range of
3600--7300\AA\ and a spectral resolution of $\sim$ 7\AA\ (FWHM). For
these observations, CCD rows were binned by a factor of 2, yielding a
final spatial sampling of 0.6 arcsec\ pixel$^{-1}$.  The observations
cover the full spectral range in a single frame which contains all
the lines of interest. Furthermore, they have sufficient spectral 
resolution to separate
[OIII]$\lambda$4363 from nearby H$\gamma$ and to distinguish between
narrow nebular and broad WR emission lines. Total exposure time was 60 
minutes and was
broken up into 3 subexposures, 20 minutes each, to allow
for more effective cosmic ray removal. All exposures were taken at
small airmasses ($\leq$1.27), so no correction was made for
atmospheric dispersion. The seeing during the observations was 1$''$ FWHM.
The slit was oriented in the direction with position
angle P.A.=--30$^\circ$ to permit observations of all stellar clusters and
low-intensity extended emission. Figure 1 shows the slit orientation 
superposed on the $R$ image of SBS 0335--052.
The spectrophotometric standard star PG 0216+032 was observed for
flux calibration. Spectra of He--Ne--Ar comparison lamps were obtained
before and after each observation to provide wavelength calibration.

Data reduction of spectral observations was carried out at 
NOAO headquarters in Tucson using
the IRAF\footnote {IRAF: the Image Reduction and Analysis Facility is
distributed by the National Optical Astronomy Observatories, which is
operated by the Association of Universities for Research in Astronomy,
In. (AURA) under cooperative argeement with the National Science
Foundation (NSF).} software package. This included bias subtraction,
cosmic-ray removal and flat-field correction using exposures of a
quartz incandescent lamp. 
After wavelength mapping, night sky background subtraction, and correcting
for atmospheric extinction, each frame was calibrated to absolute fluxes.
One-dimensional spectra were extracted by
summing, without weighting, of different numbers of rows along the slit 
depending on the exact region of interest.  
The spectrum of SBS 0335--052 in the region 3700\AA \ $\leq
\lambda \leq $ 7300\AA \ is shown in Figure 2. The continuum was fitted
after removal of the emission lines, and
line intensities were measured by fitting Gaussians to the profiles.

We have adopted an iterative procedure to derive both
the extinction coefficient C(H$\beta$) and the absorption equivalent
width for the hydrogen lines simultaneously from the equation (Izotov, Thuan \&
Lipovetsky 1994):

%
%
\begin{equation}
\frac{I(\lambda)}{I(H\beta)} = \frac{EW_e(\lambda)+EW_a(\lambda)}{EW_e(\lambda)} 
\frac{EW_e(H\beta)}{EW_e(H\beta)+EW_a(H\beta)} \frac{F(\lambda)}{F(H\beta)}  
10^{[C(H\beta)f(\lambda)]},  
\end{equation}
where $I$($\lambda$) is the intrinsic line flux and $F$($\lambda$) is the
observed line flux corrected for atmospheric extinction.
$EW$$_e$($\lambda$) and $EW$$_a$($\lambda$) are the equivalent widths of
the observed emission line and the underlying absorption line,
respectively, and $f$($\lambda$) is the reddening function, normalized at
H$\beta$, which we take from Whitford (1958).

We used the theoretical ratios from Brocklehurst (1971) at the electron
temperature estimated from the observed [OIII]($\lambda$4959 +
$\lambda$5007)/$\lambda$4363 ratio for the intrinsic hydrogen line
intensity ratios. For lines other than  hydrogen $EW$$_a$($\lambda$)=0 so
Eq. (1) reduces to
 
\begin{equation}
\frac{I(\lambda)}{I(H\beta)} = \frac{F(\lambda)}{F(H\beta)} 10^{[C(H\beta)f(\lambda)]}.
\end{equation}

The observed and corrected line intensities, extinction coefficient, 
and equivalent widths of
the stellar hydrogen absorption lines are given in Table 1 along with
the uncorrected H$\beta$ flux and H$\beta$ equivalent width for the brightest
part of SBS 0335--052 with an aperture of 1$''$$\times$6$''$.

\section{HEAVY ELEMENTS AND HELIUM ABUNDANCES}

Since SBS 0335--052 is the second most metal-deficient BCG known after
I Zw 18, it is of special interest to determine the heavy element and helium 
abundance in this very low-metallicity environment. SBS 0335--052 is a
chemically unevolved galaxy, so we can expect that the amount of helium
synthesized by stars is small and that the helium mass fraction in 
its HII region is close to the primordial value. As we see from Table 1, 
emission lines of several heavy elements have been detected in
SBS 0335--052. This gives us the
opportunity to derive heavy element abundance ratios and to place constraints on
low-metallicity stellar nucleosynthesis and chemical evolution models.
To derive element abundances we follow the procedure described in
detail by Izotov, Thuan \& Lipovetsky (1994). It is known that the electron
temperature, $T_e$, is different in high- and low-ionization zones of HII
regions (Stasinska 1990), and we have chosen to determine $T_e$(OIII) from the
[OIII]$\lambda$4363/($\lambda$4959+$\lambda$5007) ratio and $N_e$(SII) from
the [SII]$\lambda$$\lambda$6717/$\lambda$6731 ratio. 
We adopt $T_e$(OIII) for the 
derivation of He$^+$, He$^{2+}$, O$^{2+}$, Ne$^{2+}$ and Ar$^{3+}$ ionic 
abundances. To derive the electron temperature for the O$^+$ ion, we have used 
the relation between
$T_e$(OII) and $T_e$(OIII) (Izotov, Thuan \& Lipovetsky 1994), based on the
photoionization models of Stasinska (1990). $T_e$(OII) has been used to
derive the O$^+$, N$^+$ and Fe$^+$ ionic abundances. For Ar$^{2+}$
and S$^{2+}$ we have used an electron temperature intermediate between
$T_e$(OIII) and $T_e$(OII) following the prescriptions of Garnett (1992).
Total element abundances have been derived after correction for unseen stages
of ionization as described by Izotov, Thuan \& Lipovetsky (1994).
 
\subsection{Heavy Element Abundances in the Central Part of SBS 0335--052}

In Table 2 we show the adopted electron temperatures for different ions and the
electron number density for the central 1$''$$\times$6$''$ portion of SBS 
0335--052, along with ionic abundances, ionization correction factors
and total heavy element abundances. It is well established theoretically that
the oxygen seen in HII regions is a primary element produced by massive stars
with M$\geq$10M$_\odot$. Other $\alpha$-process elements seen in the
spectra of HII regions, such as neon, argon, and sulfur, are generally
thought to be primary as well. The Ne/O, S/O and Ar/O ratios are independent of
O/H (Vigroux, Stasinska \& Comte 1987; Garnett 1989; Thuan, Izotov \& 
Lipovetsky 1995; Izotov, Thuan \& Lipovetsky 1996). 

The situation for
nitrogen is more complex. While in spiral galaxies spectral observations show
that the N/O ratio increases with O/H, in low-metallicity galaxies N/O
is found to be constant and independent of O/H (Lequeux et al. 1979; Kunth \&
Sargent 1983; Campbell, Terlevich \& Melnick 1981; Thuan, Izotov \& Lipovetsky
1995). Because SBS 0335--052 so metal-deficient
it is important to examine whether the trends in heavy element abundance ratios 
continue to the extremely low oxygen abundances found in I Zw 18 and SBS 
0335--052.
For SBS 0335--052 we derive an oxygen abundance, 12 + log (O/H) = 7.33$\pm$0.01,
somewhat larger
than value 7.30 derived by Melnick, Heydari-Malayeri \& Leisy (1992).
The value of 12 + log (O/H) = 7.0 -- 7.1 found by
Izotov et al. (1990ab) was an underestimate caused mainly by nonlinearity of 
the detectors used and by saturation of strong emission lines.

In Figure 3 we show the N/O, Ne/O, Ar/O and S/O ratios for SBS 0335--052
(filled circles). We also show the data from Thuan, Izotov \& Lipovetsky
(1995) and Izotov, Thuan \& Lipovetsky (1996) (open circles) for a sample
of low-metallicity BCGs. We used the [NII] $\lambda$6584 emission line 
to derive the nitrogen abundance for SBS 0335--052. However, our relatively 
low spectral resolution is insufficient to resolve it from H$\alpha$ and its 
intensity was derived after subtraction of the broad H$\alpha$ wings. Therefore,
the derived nitrogen abundance is somewhat uncertain and 
higher spectral resolution
observations are needed to improve this value. As it is evident from Figure 3,
the N/O, Ar/O and S/O abundance ratios in SBS 0335--052 are in excellent 
agreement with those derived for other BCGs, and we confirm 
the important conclusion
by Thuan, Izotov \& Lipovetsky (1995) that all these elements in 
low-metallicity BCGs are primary elements produced by the same massive stars. 

This result is especially important for nitrogen. 
While our data show that nitrogen in low-metallicity BCGs is a primary
element, Pettini et al. (1994) found very low N/O ratio in low-metallicity
high-redshift absorbing clouds which is at variance with the data for BCGs,
suggesting the presence of a significant amount of nitrogen, produced as a
secondary element. Timmes et al. (1995) have noted 
that no primary nitrogen is produced in the standard massive star models.
However, if numerical parameters governing convective overshoot are
enlarged, primary nitrogen is produced in all low metallicity massive
stars with M$>$30M$_\odot$. The synthesis of primary
nitrogen in these low metallicity massive stars occurs as the convective
helium burning shell penetrates into the hydrogen shell with violent, almost
explosive consequences. On the other hand, in massive stars with solar 
metallicity no primary nitrogen is produced even with enhanced overshoot. Thus,
chemical evolution models with primary nitrogen production in 
low-metallicity massive stars can explain the observed lack of secondary 
nitrogen in BCGs and the small dispersion in log(N/O). 

We have detected the [FeIII]$\lambda$4658 emission line in the spectrum of SBS 
0335--052 and have derived the iron abundance in HII region. Iron is a primary 
element produced during explosive nucleosynthesis
either in Type I supernovae (SN I's) with low-mass progenitors or in Type II
supernovae (SN II's) with more massive progenitors (Weaver \& Woosley 1993;
Woosley \& Weaver 1995). Due to the difference in evolution timescales of
Type I and Type II supernova progenitors, SN I's begin to contribute only when
the age of the system is $\geq$10$^9$yr. Therefore, the Fe/O ratio 
provides an important constraint on galaxy chemical evolution models. However, 
there remain 
uncertainties concerning models of explosive nucleosynthesis due to the
uncertainties in central collapsing core parameters and the 
initial conditions for
the shock wave in the supernova progenitor. It was shown by Thuan, Izotov \&
Lipovetsky (1995) that O/Fe in BCGs is overabundant by $\sim$0.34 dex, 
the same overabundance found in galactic halo stars. This implies a similar 
chemical enrichment history for BCGs and the galactic halo. The comparison
of the observed Fe/O abundance ratio and theoretical Fe/O yield 
ratio in BCGs favors massive star
explosive nucleosynthesis models with reduced iron production.
We show in Figure 4 [O/Fe] vs. [Fe/H] for the BCGs in the sample from 
Thuan, Izotov \& Lipovetsky (1995) and Izotov, Thuan \& Lipovetsky (1996) (open
circles) along with [O/Fe] for SBS 0335--052 (filled circles), derived in this
paper.
For the solar abundance we adopt 12 + log(Fe/H)$_\odot$ = 7.51 
consistent with the
meteoritic value (Holweger et al. 1991; Bi\'{e}mont et al. 1991; Hannaford
et al. 1992).
For comparison, we also plot the data for galactic disk and halo stars 
(points) from Edvardsson et al. (1993), Barbuy (1988) and Barbuy \&
Erdelyi-Mendes (1989). While other BCGs show significant overabundance of
oxygen relative to iron, the O/Fe abundance ratio in SBS 0335--052 is close
to solar. The reason for such difference is not clear. We can only propose
several possibilities: a) the contribution of iron from type I
SNe is large in SBS 0335--052 due to the presence of a significant 
(but undetected) stellar population with age $\geq$10$^9$yr;
however, [FeIII] lines are detected in the region of present star
formation and, therefore, we do not believe the significant iron yield from
SN I's;
b) [FeIII]$\lambda$4658 emission line
in the spectrum of SBS 0335--052 is contaminated by Wolf-Rayet CIII 
$\lambda$4658 emission. However, the $\lambda$4658 line is narrow, atypical of
WR lines, and 
we do not detect any other evidence for WR features in the spectrum of
SBS 0335--052; c) iron in other BCGs is significantly depleted due to the
dust formation; d) uncertainties in the ionization correction factor (ICF) for 
iron.
ICF(Fe) is large (Table 2) and dependent on the O/O$^+$ ionic ratio, where
O=O$^+$ + O$^{2+}$. Therefore, uncertainties in O$^+$ abundance could 
significantly influence iron abundance. However, this explanation is implausible
because ICF(N) also scales with O/O$^+$, whereas N/O in SBS 0335--052 is normal.

\subsection{Helium Abundance}

The extremely low oxygen abundance in SBS 0335--052 and high surface brightness 
of its HII region implies that this galaxy is one of the best objects for
determining the helium abundance at low metallicity. The high S/N
ratio spectrum obtained for SBS 0335--052 (Figure 2) allows us to measure
helium line intensities with good accuracy. The HeI line intensities
corrected for interstellar extinction are shown in the Table 1. 
For the helium abundance determination we use the HeI $\lambda$4471, 
$\lambda$5876 and $\lambda$6678 emission lines. However, HeI line intensities 
deviate from the pure recombination values  
(Izotov, Thuan \& Lipovetsky 1996) and must be corrected.
The main mechanism changing the HeI line intensities from their recombination
values is collisional excitation from the 
metastable 2$^3$S level. This mechanism depends on electron temperature and 
electron number density and is significant in SBS 0335--052 because of the high 
electron temperature $T_e$=19200K in its HII region. The line most sensitive to 
collisional enhancement in optical range is HeI $\lambda$7065.
Another mechanism which could drive HeI line intensities from their
recombination values is self-absorption in some optically thick emission lines,
such as HeI $\lambda$3889. The emission lines most sensitive to this 
fluorescence mechanism are HeI $\lambda$3889 and $\lambda$7065. 
In contrast to collisional enhancement, which increases the 
intensities of all HeI lines, the fluorescence mechanism works in such a way as
to decrease the intensity of the HeI $\lambda$3889 line as its
optical depth increases, while increasing the intensities of other lines of
interest (HeI $\lambda$4471, $\lambda$5876, $\lambda$6678 and $\lambda$7065).

In SBS 0335--052 such fluorescent enhancement could be important. Since
HeI $\lambda$3889 is blended with H8 $\lambda$3889, we have
subtracted the latter, assuming its intensity to be equal to 0.106 I(H$\beta$)
(Aller 1984). Then the intensity of HeI $\lambda$3889 corrected 
for interstellar extinction is equal to 0.065 I(H$\beta$), 
which gives an observationally inferred ratio
I($\lambda$3889)/I($\lambda$5876) = 0.65 compared to the 
recombination value of 1.08 at $T_e$ = 20000K (Brocklehurst 1972). 
Taking into account that collisional enhancement factors for HeI $\lambda$3889
and $\lambda$5876 emission lines are nearly the same 
(Kingdon \& Ferland 1995) we conclude
that the intensity of HeI $\lambda$3889 is reduced due to 
self-absorption. To correct HeI emission line intensities for collisional and 
fluorescent enhancement we follow the approach described by Izotov, Thuan \&
Lipovetsky (1996). We assume that the electron temperature in He$^+$ zone is 
equal to that in the O$^{2+}$ zone. However, we do not use the electron
number density derived from [SII] $\lambda$6717/$\lambda$6731 for two reasons:
1) the S$^+$ and He$^+$ zones not coincide; 
2) although the electron number density $N_e$(SII)=393 cm$^{-3}$ is definitely 
larger than the low density limit for the [SII]
line ratio, we find that the width of [SII]$\lambda$6731 for unknown reasons 
is 9\% larger than that of [SII]$\lambda$6717. Therefore, rather than using
$N_e$(SII), we take into consideration 5 HeI emission lines: 
$\lambda$3889, $\lambda$4471, $\lambda$5876, $\lambda$6678 and $\lambda$7065
and solve problem self-consistently to reproduce the theoretical 
HeI line intensity
recombination ratios for two unknown quantities: 1) the electron number
density in the He$^+$ zone and 2) the optical depth $\tau$($\lambda$3889) in
HeI $\lambda$3889 line.

The best solution is found for electron number density $N_e$(He$^+$)=
153 cm$^{-3}$ and optical depth $\tau$($\lambda$3889)=1.5. 
The helium abundance for individual lines along with
correction factors for collisional and fluorescent enhancement are given in
Table 3. The helium mass fraction in SBS 0335--052, Y=0.245$\pm$0.006, is in
close agreement with that derived for other low-metallicity 
BCGs and is close to the
primordial helium mass fraction Y$_p$=0.243$\pm$0.003 derived by Izotov,
Thuan \& Lipovetsky (1996).

\subsection{Oxygen Abundance Distribution}

The presence of large number of massive stars in the 
central part of SBS 0335--052
implies that the enrichment of ionized gas with heavy elements occured in the 
short
time scale of $\sim$ 10$^6$--10$^7$yr, comparable to the lifetime 
of massive stars. The scenario of self-enriched giant HII regions 
has been proposed by
Kunth \& Sargent (1986), who suggest that new heavy element ejecta originating
from stellar winds and supernovae of type II initially mix exclusively with the
ionized gas in the HII zone, waiting for further mixing with the cold gas
during the long interburst phase. Roy \& Kunth (1995) have analyzed different
mechanisms of interstellar medium mixing and found that ionized gas is well
mixed due to Rayleigh-Taylor and Kelvin-Helmholz instabilities 
on time scales 1.5$\times$10$^6$ yr within regions of 100 pc size. Martin (1996)
has studied the oxygen abundance distribution inside the central 530 pc of the
galaxy I Zw 18, where auroral [OIII] $\lambda$4363 is observed and has 
found that oxygen abundance within the central 530 pc is nearly 
constant  and is within 20\% that of the 
NW HII region (12 + log (O/H) = 7.1 -- 7.3). She argued that detection of 
a superbubble establishes
a timescale ($\sim$15--27 Myr) and spatial scale ($\sim$900 pc) for dispersing
the recently synthesized elements. 

The [OIII]$\lambda$4363 emission line in SBS 0335--052 is observed within
the region of 3.6 kpc. This allows us to measure oxygen abundance and
to study processes of interstellar medium mixing in a region 7 times larger
than in I Zw 18. In Figure 5a we show intensities of [OIII]$\lambda$4363 and
$\lambda$5007 relative to H$\beta$, corrected for interstellar extinction, and 
the distribution of the interstellar extinction coefficient, C(H$\beta$),
in the central 3 kpc region of SBS 0335--052 in the direction NW -- SE.
The origin is taken with respect to the maximum of the continuum flux near the 
H$\beta$ emission 
line. Six blue stellar clusters detected by Thuan, Izotov \& Lipovetsky (1996),
are all located in the region from --250 pc to 250 pc. We find that 
interstellar 
extinction is significant in the central 3 arcsec or $\sim$ 800 pc of the
galaxy. Sites of enhanced absorption coincide spatially with the red region
detected by Thuan, Izotov \& Lipovetsky (1996). We argue that their assumption
is correct; the red color is caused by presence of
dust, however, we find somewhat lower values
for extinction coefficient, C(H$\beta$) = 0.2 -- 0.3, (cf. the value
0.56 derived by Melnick, Heydari-Malayeri \& Leisy (1992) for knot A). The
maximal value of C(H$\beta$)=0.27 is equivalent to a reddening of 
$E(B-V)$ = 0.18.
In Figure 5b we show the distribution of electron temperature $T_e$ (solid line)
and oxygen abundance 12+log (O/H) (dashed line) in central 3 kpc of SBS 0335--052.
For comparison, the oxygen abundance distribution in I Zw 18 is shown by the
dotted line. In the brightest 1 kpc of SBS 0335--052, the temperature is
nearly constant and lies in the range $T_e$= 18500 -- 20000 K. In this region,
the oxygen abundance is also nearly constant and lies in the very narrow 
range 12 + log (O/H) = 7.30 -- 7.35. This constancy of the oxygen abundance 
implies that the mixing of oxygen-rich supernova remnants with interstellar
medium happened very quickly, $\sim$10$^6$ yr. The distribution of the
electron temperature in NW direction is characterized by several 
jumps. The dispersion
of $T_e$ is $\sim$5000 K, caused mainly by observational uncertainties
due to the low intensities of emission lines at the distance greater than
1 kpc from the origin. However, it is probable, that some part of the dispersion
in $T_e$ could be explained by propagation of shocks in NW direction. Indeed,
HST images of SBS 0335--052 clearly show elongated structure and several
superbubbles in that direction (Thuan, Izotov \& Lipovetsky 1996). 
Some dispersion in oxygen abundance in the NW direction could be in 
part due to the
incomplete mixing at scales larger than 1 kpc. We note the
small gradient of oxygen abundance with 12 + log (O/H) $\sim$ 
7.1 at a distance of 2 kpc, which is 1.5 times lower than value at the origin.
The size of HII region in I Zw 18 is several times smaller, but the oxygen
abundance is nearly in the same range, 7.1 -- 7.3, as in SBS 0335--052.
Therefore, the oxygen abundance distribution in I Zw 18 and SBS 0335--052 is
similar, the main difference is caused by different scales -- in SBS 0335--052
the population of massive O-stars is one order of magnitude 
larger than in I Zw 18.
The small gradient in oxygen abundance found in SBS 0335--052 on scales $\sim$
3 kpc is evidence in favor of fast mixing of ionized gas 
due to turbulent motions produced by fast shocks, moving in the direction 
SE -- NW.  However we should adopt the characteristic
turbulent velocity of $\sim$10$^3$ km s$^{-1}$ for efficient mixing on  
these scales.  
Velocities of order 10$^3$ km s$^{-1}$ have been detected by Roy et al. 
(1992), Skillman \& Kennicutt (1993) and Izotov et al. (1996) in several
dwarf galaxies. In \S6 we discuss the existence of such fast gas motion
in SBS 0335--052.


\section{THE NATURE OF HEII $\lambda$4686 EMISSION}

   The strong nebular HeII $\lambda$4686 emission line has been detected in SBS 
0335--052 (Figures 2, 8). Its intensity, $\sim$3\% of H$\beta$,
is close to that observed in I Zw 18 (Skillman \& Kennicutt 1993) and in some
other low-metallicity BCGs (Campbell, Terlevich \& Melnick 1986;
Terlevich et al. 1991; Izotov, Thuan \& Lipovetsky 1994, 1996; Izotov et
al. 1996) and is several orders of magnitude larger than theoretical values
predicted by models of photoionized HII regions. It was suggested by
Bergeron (1977), that the HeII emission in dwarf emission-line galaxies could
arise in the atmospheres of Of stars. Garnett et al. (1991) 
presented observations of nebulae in nearby dwarf galaxies with strong narrow
HeII $\lambda$4686 emission lines and examined several possible 
excitation mechanisms, concluding
that the radiation field associated with star-forming regions can be 
harder than previously suspected. The hottest main sequence stars
have temperatures not exceeding values 60000K (Campbell 1988); plane-parallel
non-LTE atmosphere models for such stars under-produce the required number
of He$^+$ ionizing photons by roughly four orders of magnitude (Garnett et al.
1991). On the other hand, models for massive star evolution which include mass
loss predict that the most massive stars evolve blueward to larger 
effective temperatures and become Wolf-Rayet stars.
Schaerer \& de Koter (1996) calculated non-LTE atmosphere models
taking into account line blanketing and stellar winds and found that the flux
in the HeII continuum is increased by 2 to 3 orders of magnitudes compared to
predictions from plane-parallel non-LTE model atmospheres and 3 to 6 orders
of magnitudes compared to predictions from plane parallel LTE model atmospheres.
Including these predictions, Schaerer (1996) synthesized the nebular and 
Wolf-Rayet HeII $\lambda$4686 emission in young starbursts. For
metallicities 1/5Z$_\odot$$\leq$Z$\leq$Z$_\odot$, he predicted a strong nebular 
HeII emission due to a significant fraction of WC stars in early WR phases of
the burst. His predictions of the nebular HeII (typically I(HeII)/I(H$\beta$) 
$\sim$ 0.01 -- 0.025) agree well with the 
observations in Wolf-Rayet galaxies. However, this mechanism is not appropriate
for SBS 0335--052 and other low-metallicity BCGs due to the fact that at low 
metallicities the efficiency of stellar winds is low, 
as suggested by the lack of
the broad WR emission lines in their spectra (Figures 2, 8). It was suggested
by Garnett et al. (1991) that radiative shocks in giant HII regions can
produce relatively strong HeII emission under 
certain conditions. The strength of
the HeII emission is sensitive mostly to the velocity of the shock, reaching
maximum for $V$$_{shock}$$\sim$120 km s$^{-1}$, dropping rapidly at higher
velocities. The third
mechanism, discussed by Garnett et al. (1991) is photoionization of HII region
by X-rays produced by massive X-ray binary stars.

In Figure 6 we show spatial distribution of continuum and emission line 
intensities for different emission lines in SBS 0335--052. We note that 
distribution of continuum intensity
measured near the H$\beta$ emission line is shifted by $\sim$200 pc 
to NW relative to the spatial distribution of emission lines intensities. 
The width of continuum 
distribution at half maximum is $\sim$ 700 pc, close to the diameter
of region in SBS 0335--052 where six blue compact clusters are located (Thuan,
Izotov \& Lipovetsky 1996). Therefore, the 
optical continuum in the central part 
of SBS 0335--052 is produced by stars, mainly O and B. 
Thuan, Izotov \& Lipovetsky (1996) found a $(V-I)$ 
color gradient of clusters in 
the direction SE--NW with bluest cluster at the SE edge. They explain the color 
gradient mainly by reddening due to the presence of dust patches.
However, some evolutionary effect could be present as follows from Figure 6: the
maximum in the emission line intensities coincides with location of bluest and
youngest stellar cluster. Other clusters probably have 
somewhat larger ages;
most massive stars have moved away from the main sequence. This conclusion is 
confirmed by the equivalent widths of the H$\beta$ emission line:
158\AA\ at the origin, where the intensity of continuum is largest, and to
250\AA\ at the maximum in the emission line intensity distribution. 
Consequently, ionizing
photon fluxes from the older clusters are smaller.
The only exception among nebular emission lines is HeII 
$\lambda$4686 emission line. Its spatial distribution
is nearly the same as stellar continuum distribution. Therefore, one of
the probable mechanisms of HeII emission is hard emission from hot stars which
are now at post-main-sequence stage. We exclude the 
possibility that this emission
is produced due to ionization of He$^+$ by radiation of O stars on the main 
sequence. If this were the case, the HeII intensity spatial distribution would 
be coincident with the spatial distribution of other emission lines.
Due to the fact that we haven't detected WR stars in SBS 0335--052, 
He$^+$ ionization could be produced by massive X-ray binaries. We can 
estimate the number of massive X-ray binary systems necessary to produce 
observed luminosity of L(HeII 
$\lambda$4686) = 5.74$\times$10$^{38}$ erg s$^{-1}$. Scaling directly from the
observed HeII luminosity 1.5$\times$10$^{35}$ erg s$^{-1}$ of the nebula 
surrounding LMC X-1 (Pakull \& Angebault 1986), the HeII luminosity
in SBS 0335--052 implies the presence of $\sim$4000 massive X-ray binary 
systems. This value is in agreement with the equivalent number of O7-stars 
N(O7)=5000 inferred from the luminosity of H$\beta$ emission line
L(H$\beta$) = 2.06$\times$10$^{40}$erg s$^{-1}$. Additionally, it is follows
from Figure 5a, that the origin of HeII $\lambda$4686 in fast shocks is not
ruled out due to the fact that the intensity of this line, 0.04 -- 0.05 
relative to H$\beta$, is observed in the NW direction at the distance 
$\sim$ 1 kpc, beyond the stellar clusters.

\section{KINEMATICS OF IONIZED GAS}

The presence of filaments and arcs in SBS 0335--052 found by Thuan, Izotov 
\& Lipovetsky (1996) from HST WFPC2 images implies fast moving gas 
and complex dynamics of the ionized gas in HII region due to the supernovae 
activity. In Figure 7a we show the velocity distribution of ionized gas from
H$\beta$ $\lambda$4861, [OIII]$\lambda$5007 and H$\alpha$ $\lambda$6563 
emission lines. For comparison, the night sky [OI]$\lambda$5577 is also shown. 
The velocity distribution
is similar for all nebular lines. However, contrary to 
expectations from imaging,
the velocity dispersion of ionized gas is very small $\leq$10 km s$^{-1}$ in 
the inner part of HII region. We note a small velocity gradient in the NW 
direction from the brightest part of the galaxy, 
$\sim$ 20 km s$^{-1}$kpc$^{-1}$. The radial velocity of the central part of the 
galaxy is in good agreement with the
radial velocity, derived from VLA HI $\lambda$21 cm observations
(Thuan et al. 1996). At large distances
from the HII region center, the radial velocity is slightly increased by 
$\sim$30--40 km s$^{-1}$. Combining the presence of gaseous filaments
with the absence of velocity gradients, we conclude from the radial velocity 
distribution 
that the ionized gas motion occurs in the direction perpendicular to the 
line-of-sight
and perpendicular to the edge-on HI gas cloud. Such geometric orientation
of the galaxy and large column density of neutral hydrogen in the direction of 
the stellar clusters could be the reason why Ly$\alpha$ emission was not 
detected in HST GHRS spectrum by Thuan, Izotov \& Lipovetsky (1996). Following
Charlot \& Fall (1993), even if the dust is absent in the gas, 
the number of Ly$\alpha$
photons which escape in the plane of edge-on neutral gas cloud is significantly
lower than in the direction normal to the gaseous disk plane.

In Figure 7b the full width at half maximum (FWHM) distribution for brightest
lines is shown. For comparison, the FWHM for night sky line 
[OI]$\lambda$5577, which does not show changes along the slit, is also 
presented. We note that FWHM of the nebular lines 
increases from SE to NW. If real, this effect could be explained by
transformation of ordered motion of supernova shells to chaotic turbulent
motion due to instabilities in ionized gas discussed by Roy \& Kunth (1995).
We can estimate the characteristic FWHM caused by turbulent motion 
at the NW edge of the galaxy from relation FWHM$_{tur}^2$=FWHM$_{tot}^2$ --
FWHM$_{inst}^2$, where FWHM$_{tot}$ and FWHM$_{inst}$ are total and 
instrumental FWHM, the latter being derived from night sky 
[OI]$\lambda$5577 line.
The FWHM$_{tur}$ derived from [OIII]$\lambda$5007 at the NW edge of HII
region is equivalent to a turbulent velocity of
$\sim$ 50 to 100 km s$^{-1}$. This turbulent velocity can 
explain effective mixing of ionized gas on scales of 1 kpc during 
$\sim$10$^7$yr and nearly constant
oxygen abundance. 

In Figure 8 we show the part of the spectrum with H$\beta$ and [OIII]$\lambda$
4959, 5007 emission lines from the central part of the galaxy. The 
H$\beta$ emission 
line has broad wings with FWZI$\sim$50\AA; we attribute these to
fast moving ($\sim$1000 km s$^{-1}$) supernovae remnants which, at larger 
distances, transform to random motions. We suggest that it is the only
mechanism for ionized gas mixing at scales of several kpc. The broad wings of
[OIII]$\lambda$4959, 5007 emission lines are masked by nearby faint lines and
not so evident, but they are possibly present. Roy et al. (1992) and
Izotov et al. (1996) have shown that the broad H$\beta$ wings 
in the low-metallicity BCGs are associated with broad wings in 
[OIII]$\lambda$4959, 5007 emission lines. 
 
\section{THE ORIGIN OF EXTENDED UNDERLYING LOW-INTENSITY EMISSION} 

The study of photometric and spectrophotometric properties of low-intensity
extended emission around star-forming regions in BCGs is of special interest.
The low metallicity of these galaxies implies that among them could exist young
galaxies where star formation is occuring for the first time (Searle, Sargent
\& Bagnuolo 1973; Kunth \& Sargent 1986). In the majority of BCGs the
underlying low-surface-brightness emission component has been detected,
characterized by red colors which are consistent with the 
presence of K and M stars (Loose \& Thuan 1985; 
Kunth, Maurogordato \& Vigroux 1988). These galaxies, therefore, are not young 
and have already experienced several episodes of 
star formation. It is believed that the best young galaxy 
candidates are I Zw 18 and SBS 0335--052. 
Hunter \& Thronson (1995) found, on the
basis of HST images for I Zw 18, that the colors of the underlying diffuse
component are consistent with those of B or A stars, with no evidence for
long-lived red stars. Thuan, Izotov \& Lipovetsky (1996) obtained $V$ and $I$
images for SBS 0335--052 and detected a low-intensity extended component
$\sim$14$''$ in diameter elongated in the SE--NW direction. 
The color $(V-I)$$_0$
for this component at large distance ($r$$>$2$''$ or 
520 pc in linear scale) from 
the central part of the galaxy where stellar clusters reside is  
0.0$\leq$$(V-I)$$_0$$\leq$0.2 which is characteristic of A stars. 
Several gaseous filaments and arcs are superimposed on this underlying 
component, which led Thuan, Izotov \&
Lipovetsky (1996) to suggest that the underlying component is gaseous in
nature and SBS 0335--052 is a young galaxy, probably 
undergoing its very first burst of star formation. 

To examine the nature of the underlying diffuse component, Lipovetsky et al. (1996)
obtained $R$ and $I$ images with the 3.5 meter Calar Alto telescope. We compare
the total $I$ magnitudes and $I$ brightness distributions 
for SBS 0335--052 obtained
with HST and the 3.5 meter Calar Alto telescope and find them to be in 
good agreement. The
difference in total $I$ magnitude is only 0.05 mag inside the 25 
mag/arcsec$^2$ isophote. Its rather blue $(R-I)$ color lies in the range 
--0.6 -- 0.0 mag (Lipovetsky et al. 1996). From models of stellar population 
synthesis  Leitherer \& Heckman (1995) have derived $(V-I)$ = 0.0 and
$(R-I)$ = 0.0 for an instantaneous burst at the age log $t$ = 6.5 and 
0.60 and 0.30 at the age log $t$ = 8.0, where time $t$ is in yr. 
Therefore, both $(V-I)$ and $(R-I)$ colors in SBS 0335--052 are 
inconsistent with a pure stellar origin for the diffuse component.

    Stellar emission in SBS 0335--052 is contaminated by strong nebular 
emission lines and gaseous continuous emission. In the $V$ band, 
significant flux 
arises from [OIII]$\lambda$5007, which has an equivalent width $\sim$500\AA; 
in the $R$ band H$\alpha$, with equivalent width $\sim$1000\AA, dominates. 
Only the $I$ band is free of strong emission lines, but there free-free 
and bound-free emission could be important. The emission lines are observed in 
the region outside the central 4 kpc. We now turn our attention to 
the results of broad-band photometry and ask whether the $(V-I)$ and $(R-I)$ 
colors in the
external parts of SBS 0335--052 can be explained solely by gaseous emission.
To model the pure gaseous color distribution one should obtain the intensity 
distribution of all lines of interest. Given the electron temperature,
the continuous (free-free, bound-free and two-photon) emission of
hydrogen-helium ionized gas is a function of only the H$\beta$ line 
intensity, helium abundance and wavelength (Aller 1984). 
We synthesize the $UBVRI$ colors of the gaseous emission, 
deriving the zeropoints 
for all filters from the spectrum of Vega (Castelli \& Kurucz 1994). We find 
that the zeropoints for the HST F569W ($V$) and F791W ($I$) filters used by 
Thuan, Izotov \& Lipovetsky (1996)
differ from the $V$ and $I$ passbands from Bessel's (1990) $V$ and $I$ by
only --0.03 and +0.01 respectively. Therefore, we can directly compare HST 
and ground-based observations.

In Table 4 we compare the observed colors of the underlying low-intensity 
component with those derived from models of pure 
dust-free gaseous emission at an electron
temperature of $T_e$=20000K. In the models we adopt the continuous
hydrogen-helium emission from Aller (1984) and two values of the observed 
equivalent widths of H$\beta$ -- 170\AA\ (model II in Table 4) and 350\AA\
(model III) which are representative of the outer envelope. We have taken into
account in calculations all strong emission lines with observed intensities 
relative to 
H$\beta$. Our calculations are in  good agreement with the observations for both
$(R-I)$ and $(V-I)$ colors, and thus we infer that $T_e$=20000K ionized gas
contributes significantly to the extended low-intensity emission.
We also calculate $(U-B)$ and $(B-V)$ colors for gaseous emission
in SBS 0335--052. These colors provide a good basis for discriminating
between different models. However, pure gaseous emission model is
complicated by the fact that the observed equivalent widths of H$\beta$ in the
extended envelope do not exceed 350\AA\, with mean value about 200\AA, 
while recombination theory (case B) at $T_e$ = 20000K gives value $\sim$ 800\AA\
(Aller 1984), or $\sim$3 times larger. 

If we assume that the difference between the observed and theoretical 
equivalent widths results from the presence of some stellar
light in the extended envelope, then $\sim$2/3 of the continuum 
near H$\beta$ could be stellar in origin. The observational data allow us
to estimate the upper limit of the stellar mass required to produce the 
extended low-intensity
emission. In Table 4 we show the predicted colors for a gaseous continuum
without emission lines at $T_e$=20000K (Model I). 
Except for $(U-B)$, these colors are in good agreement with those for an
instantaneous burst with an age log $t$ = 8.0. This means that the 
stellar-to-gaseous
continuum intensity ratio in $V$ band is the same as at H$\beta$. The apparent 
$V$ magnitude of outer envelope with radius $r$$>$3$''$ is $\sim$17.3, 
from which we infer absolute $V$ magnitude of $\sim$ --16 mag. To
explain the observed light of the extended envelope with an instantaneous burst
at log $t$ =8.0, we must invoke the conversion of 
$\sim$ 10$^7$M$_\odot$ of gas to
stars with a Salpeter IMF and a lower mass cutoff of 1M$_\odot$ (Leitherer \& 
Heckman 1995). This mass is significantly lower than the 
total mass of the galaxy,
but is comparable with the mass converted to stars in the central part 
of SBS 0335--052 during the 
present burst of star formation. Therefore, we cannot
exclude the possibility that some part of the extended low-intensity emission is
due to stars formed during a previous episode of star 
formation $\sim$ 10$^8$
yr ago. Additional evidence for an older stellar component comes from the
comparison of the H$\beta$ width with that of 
[OIII] $\lambda$5007 and the night sky 
[OI]$\lambda$5577 (Fig.7b). The width of the H$\beta$ emission line 
is less than that for the other lines
and, possibly, could be explained by the presence of underlying stellar
absorption from A stars. 
Corroborative evidence for the reality of this effect is provided by the fact
that the H$\gamma$ $\lambda$4340 emission line is also narrower than the nearby 
mercury night sky HgI $\lambda$4358.2 emission line.

Stellar population synthesis models predict that A stars 
dominate in the optical
range when stars were formed in a single burst with age $\sim$10$^8$yr.
We suggest that, since their formation, 
these stars have been dispersed by several kpc due to random motions with 
velocities of $\sim$10 km s$^{-1}$
and could explain the observed size of the region where the extended
underlying emission is observed. In order to determine which
model for the extended underlying emission is valid -- pure gaseous emission
in a 10$^7$yr old galaxy where the equivalent widths of the hydrogen emission
lines are reduced by a factor of $\sim$2--3 for an unknown reason 
or combined emission from stars and ionized gas
in a galaxy with age $\sim$ 10$^8$yr, we need to compare the
observed $(U-B)$ color with different model predictions. In the case of an older
stellar component (log $t$ = 8.0), the $(U-B)$ color would be significantly 
redder than that from pure gaseous emission (Table 4). $(U-B)$ color 
observations of SBS 0335--052 will help to 
select between these two origins of the underlying emission.

\section{SUMMARY}

We have presented high S/N MMT spectrophotometric observations 
of the extremely low-metallicity blue compact galaxy, SBS 0335--052, which 
the preponderence of evidence suggests is a nearby, 
young galaxy. In the present paper our main goal has been to
examine this hypothesis and to study the physical conditions and chemical
composition in the HII region in SBS 0335--052. 

From a detailed analysis of these data, we infer that:

   1. SBS 0335--052 is an extremely low-metallicity galaxy 
with oxygen abundance 12 + log(O/H) = 7.33$\pm$0.01 in its central 
brightest region. This value is in reasonable 
agreement with that derived
by Terlevich et al. (1992) and by Melnick, Heydari-Malayeri \& Leisy (1992) 
and is slightly greater than the value of $\sim$7.2 derived by Skillman
\& Kennicutt (1993) and by Izotov, Thuan \& Lipovetsky (1996) for I Zw 18,
most metal-deficient BCG.
   
   2. The abundance ratios log(N/O) = --1.59$\pm$0.03, log(Ne/O) = 
--0.81$\pm$0.02, log(S/O) = --1.56$\pm$0.03 and log(Ar/O) = --2.26$\pm$0.04
in the inner brightest part of SBS 0335--052 are in excellent agreement 
with mean values for low-metallicity BCGs and confirm the conclusion by Thuan, 
Izotov \& Lipovetsky (1995) that all these elements are primary elements
produced in the same massive stars during short time-scales. 
However, we find that the O/Fe ratio in
SBS 0335--052 is a factor $\sim$2--3 lower than values derived for other 
low-metallicity BCGs (Thuan, Izotov \& Lipovetsky 1995; Izotov, 
Thuan \& Lipovetsky 1996) and is close to that for the Sun. For the
moment the reason of this difference is unknown.

   3. To derive the helium abundance in SBS 0335--052 we have corrected the HeI
$\lambda$3889, $\lambda$4471, $\lambda$5876, $\lambda$6678 and $\lambda$7065 
emission lines strengths for collisional and fluorescent enhancement 
using Smits' (1996) HeI recombination emissivities and Kingdon \& 
Ferland's (1995) and Robbins' (1968) correction factors. The abundance of
doubly ionized helium derived from the HeII $\lambda$4686 emission line is also
taken into account. 
The helium mass fraction in the inner part of SBS 0335--052 is shown 
to be Y = 0.245$\pm$0.006, very close to value for primordial helium
abundance Y$_p$= 0.243$\pm$0.003 derived by Izotov, Thuan \& Lipovetsky (1996)
from the sample of 24 low-metallicity BCGs. Since the helium abundance in
I Zw 18 cannot be reliably determined owing to the peculiarity of its
HeI emission line intensities (Izotov, Thuan \& Lipovetsky 1996), 
SBS 0335--052 is now the most metal-deficient BCG which can be used to
infer the primordial helium abundance.

   4. The auroral [OIII] $\lambda$4363 emission line is detected in inner
part of SBS 0335--052' HII region with diameter 3.6 kpc which allows 
us to obtain an accurate estimate of the electron temperature and oxygen 
abundance. The diameter of the [OIII]$\lambda$4363 region in SBS 0335--052 
is $\sim$7 times larger than
in I Zw 18 implying, together with an H$\beta$ luminosity one order 
of magnitude larger number of 
massive O-stars in SBS 0335--052. We find the HII region in SBS 0335--052
to have a rather high electron temperature $T_e$$\sim$20000K,
with evidence for a small positive outward gradient.  
The oxygen abundance 12 + log(O/H) 
exhibits a small decrease with increasing radius ranging
from 7.1 to 7.3; a similar phenomenon has been reported for 
I Zw 18 (Martin 1996). The presence of a large number
of short-lived massive stars in SBS 0335--052 and the nearly constant oxygen
abundance in its inner 3.6 kpc imply that effective mixing of the ionized
gas has occured in the short time-scale of 10$^6$--10$^7$ yr. Nonetheless,
kinematical properties of the ionized gas in SBS 0335--052 show little
evidence of the fast gas motion 
needed for effective mixing on observed scales. The only
exception is the presence of marginally detected low-intensity broad component
of strong H$\beta$ emission line with FWZI$\sim$50\AA.

   5. The narrow nebular HeII $\lambda$4686 emission line is found to 
be very strong in SBS
0335--052 compared with predictions of photoionized HII region models.
The intensity of this line relative to H$\beta$ varies spatially 
and has a maximum value of 0.06 at 500 pc NW of the
brightest part of the galaxy. The large intensity of HeII $\lambda$4686
implies the presence of a strong HeII continuum beyond the 228\AA, which
is several orders of magnitude larger than values predicted by plane-parallel
LTE and non-LTE stellar atmosphere models. The observed shift in the
HeII $\lambda$4686 intensity distribution along the slit as compared with 
intensities of other nebular emission lines suggests that its origin
is not associated with
hot main-sequence O-stars. Schaerer (1996) has proposed that 
strong HeII $\lambda$4686 could be produced by Wolf-Rayet stars.
However, this would not seem to be the case for SBS 0335--052 where,
because of its very low metallicity, WR stars are unlikely to exist.
We suggest that the strong
HeII $\lambda$4686 emission line is connected in some way with evolved 
massive stars; however,
the particular mechanism (massive X-ray binaries, shocks from supernovae, and
others) is at present unknown.

   6. We use emission line intensities along the slit 
oriented in SE--NW direction to model colors $(U-B)$, $(B-V)$, $(R-I)$ 
and $(V-I)$
of pure gaseous emission at electron temperature $T_e$=20000K. Taking into
consideration free-free, bound-free, two-photon continuum emission and
\underline{observed} equivalent widths for emission lines we found good 
agreement between observed and theoretical $(R-I)$ and $(V-I)$ colors suggesting
in favor of a significant contribution of
gaseous emission in extended low-intensity envelope. However,
observed equivalent width of H$\beta$ emission line is $\sim$ 
2--3 times lower than
value expected from pure gaseous emission. Besides that, we found that $(R-I)$
and $(V-I)$ colors of gaseous continuum at $T_e$=20000K are close to those
for stellar population with age 10$^8$yr. At last, widths of H$\gamma$
and H$\beta$ lines are narrower than instrumental profiles, probably, due to
presence of underlying stellar absorption from A stars. These arguments
suggest in favor of an older stellar population with age $\sim$10$^8$yr, which
gives $\sim$2/3 of the light in the stellar continuum
outside of the inner region $\sim$500 pc in diameter 
where the young clusters reside. We estimate the mass of older
stellar population of $\sim$10$^7$M$_\odot$, which is significantly lower
(2 orders of magnitude) than the mass of neutral gas in SBS 0335--052, but
it is comparable with mass of present burst of star formation. Hence,
observational data suggest that SBS 0335--052 is young galaxy with age of
$\sim$10$^8$yr experiencing now at least
the \underline{second} short episode of star 
formation with duration $\leq$ 10$^7$yr,
or the \underline{first} episode of propagating star formation
in the direction NW--SE with duration of 10$^8$ yr, 
as suggested in part by the color gradient for stellar
clusters found by Thuan, Izotov \& Lipovetsky (1996). To confirm the presence
of an underlying stellar component, it is important to derive the $(U-B)$ color,
which is very different for gaseous and stellar emission. We also note the
importance of spectral observations in analyzing photometric data obtained in 
searches for young galaxies.

\acknowledgements

Y.I.I., V.A.L., N.G.G. and A.Y.K. have been supported by INTAS 
international grant No. 94-2285. Y.I.I., V.A.L. and A.Y.K. thank 
R.Green and the staff of
NOAO for their kind hospitality. Special thanks are due to J.Barnes for
help with the processing with IRAF.  
The authors are very grateful U.Hopp for providing the new $R$, $I$ 
data obtained with 3.5m telescope at Calar Alto. Y.I.I. thanks the
Smithsonian Institution for partial financial support of his sojourn in
Tucson.  F.H.C. and C.B.F. acknowledge the support of the National Science
Foundation under grant AST 93-20715.

\clearpage

\figcaption[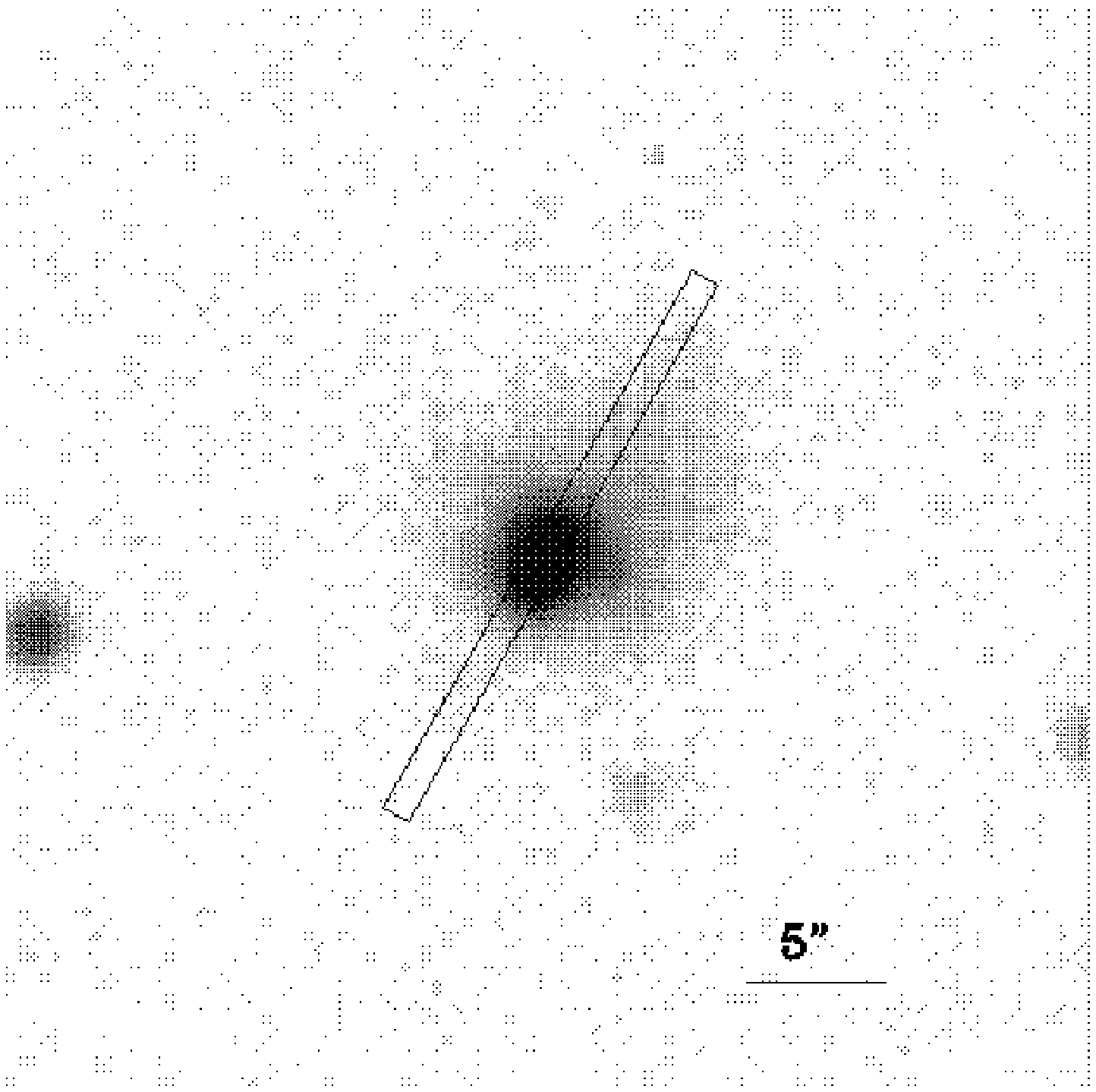]{$R$ image of SBS 0335--052, obtained with 3.5 meter Calar Alto 
telescope. North is up, East is left. The spectrum has been obtained with 
the slit oriented NW--SE along the major axis of underlying emission with
P.A. = --30 degree.} 

\figcaption[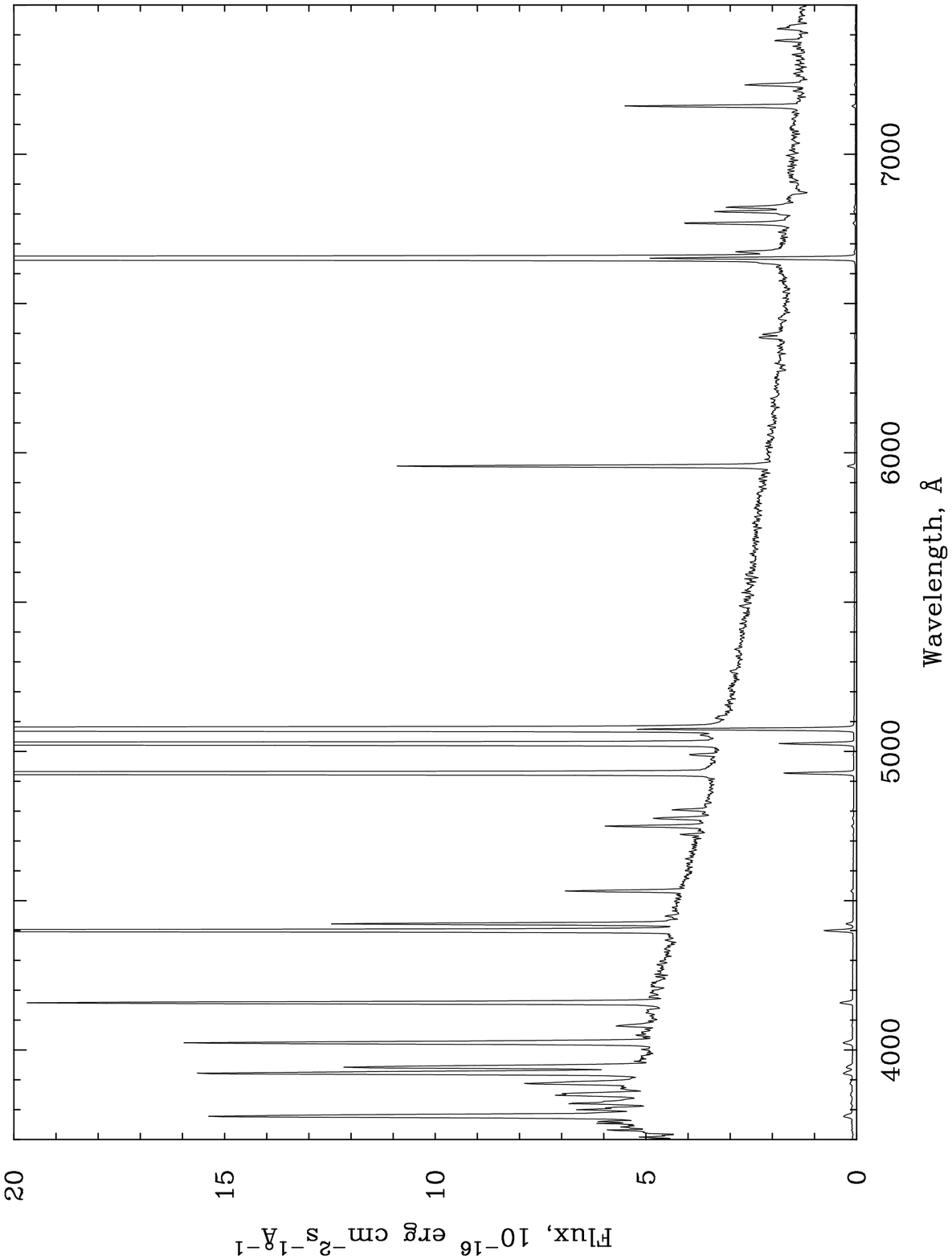]{MMT spectrum of SBS 0335--052 extracted from the brightest 
1$''$$\times$6$''$ part of the galaxy. The slit is oriented as shown in
Figure 1. The spectrum with fluxes reduced by factor 50 is also presented
to show strong emission lines.}

\figcaption[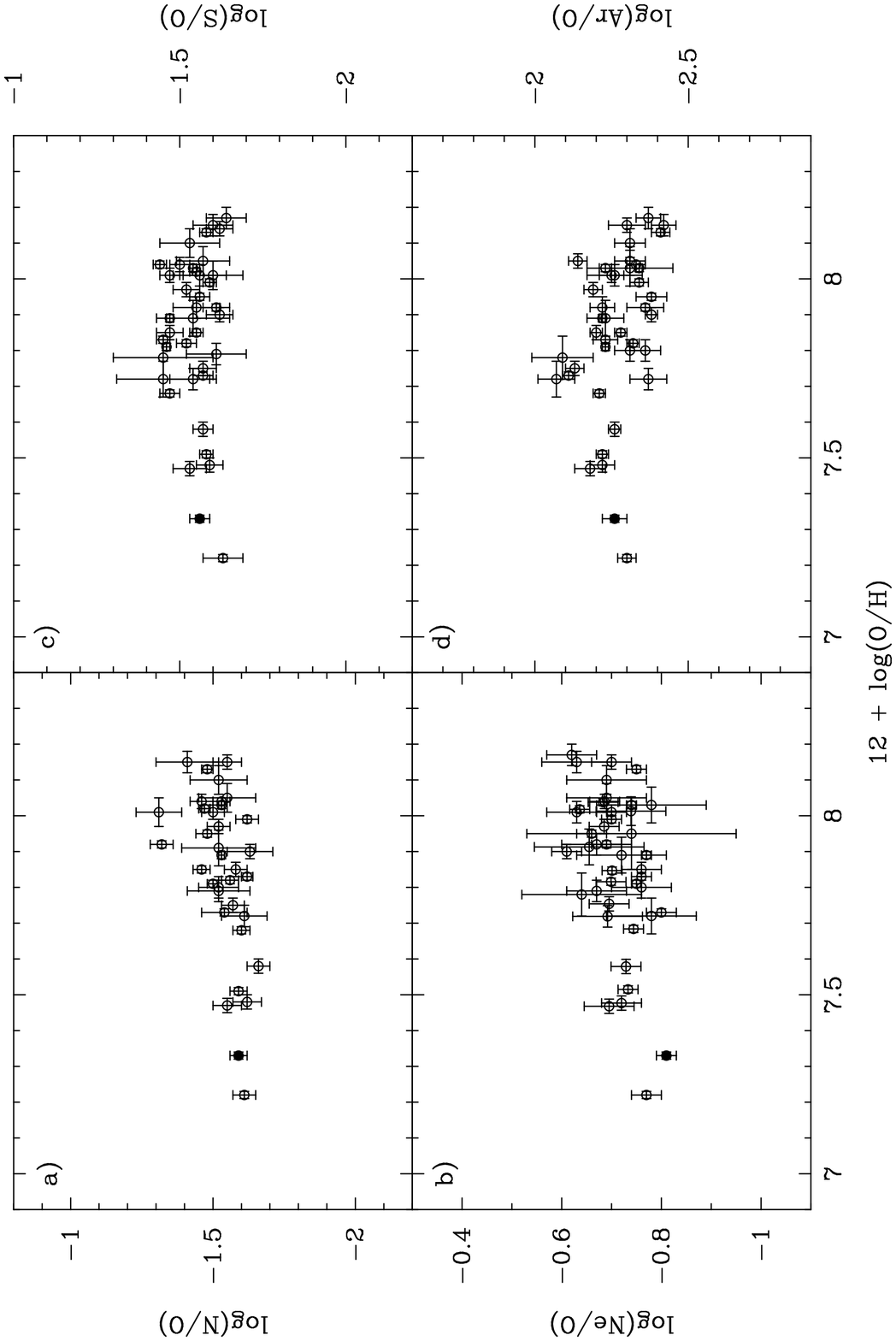]{a) Nitrogen-to-oxygen abundance ratio vs. oxygen abundance for the
sample of low metallicity blue compact galaxies (open circles) from Thuan,
Izotov \& Lipovetsky (1995) and Izotov, Thuan \& Lipovetsky (1996). The 
location of SBS 0335--052 is shown with filled circle. The most left point
is for I Zw 18. Note the low spread of points and absence of an apparent trend
in the N/O vs. O/H diagram, implying that nitrogen is a primary element produced
by massive stars. b) Location of SBS 0335--052 in the diagram of neon-to-oxygen abundance ratio vs.
oxygen abundance (filled circle). For comparison, the data for the sample
of low-metallicity BCGs (open circles) from Thuan, Izotov \& Lipovetsky (1995) 
and Izotov, Thuan \& Lipovetsky (1996) are shown.
c) Location of SBS 0335--052 in the diagram sulfur-to-oxygen abundance ratio vs.
oxygen abundance (filled circle). For comparison, the data for the sample
of low-metallicity BCGs (open circles) from Thuan, Izotov \& Lipovetsky (1995) 
and Izotov, Thuan \& Lipovetsky (1996) are shown.
d) Location of SBS 0335--052 in the diagram argon-to-oxygen abundance ratio vs.
oxygen abundance (filled circle). For comparison, the data for the sample
of low-metallicity BCGs (open circles) from Thuan, Izotov \& Lipovetsky (1995) 
and Izotov, Thuan \& Lipovetsky (1996) are shown.}

\figcaption[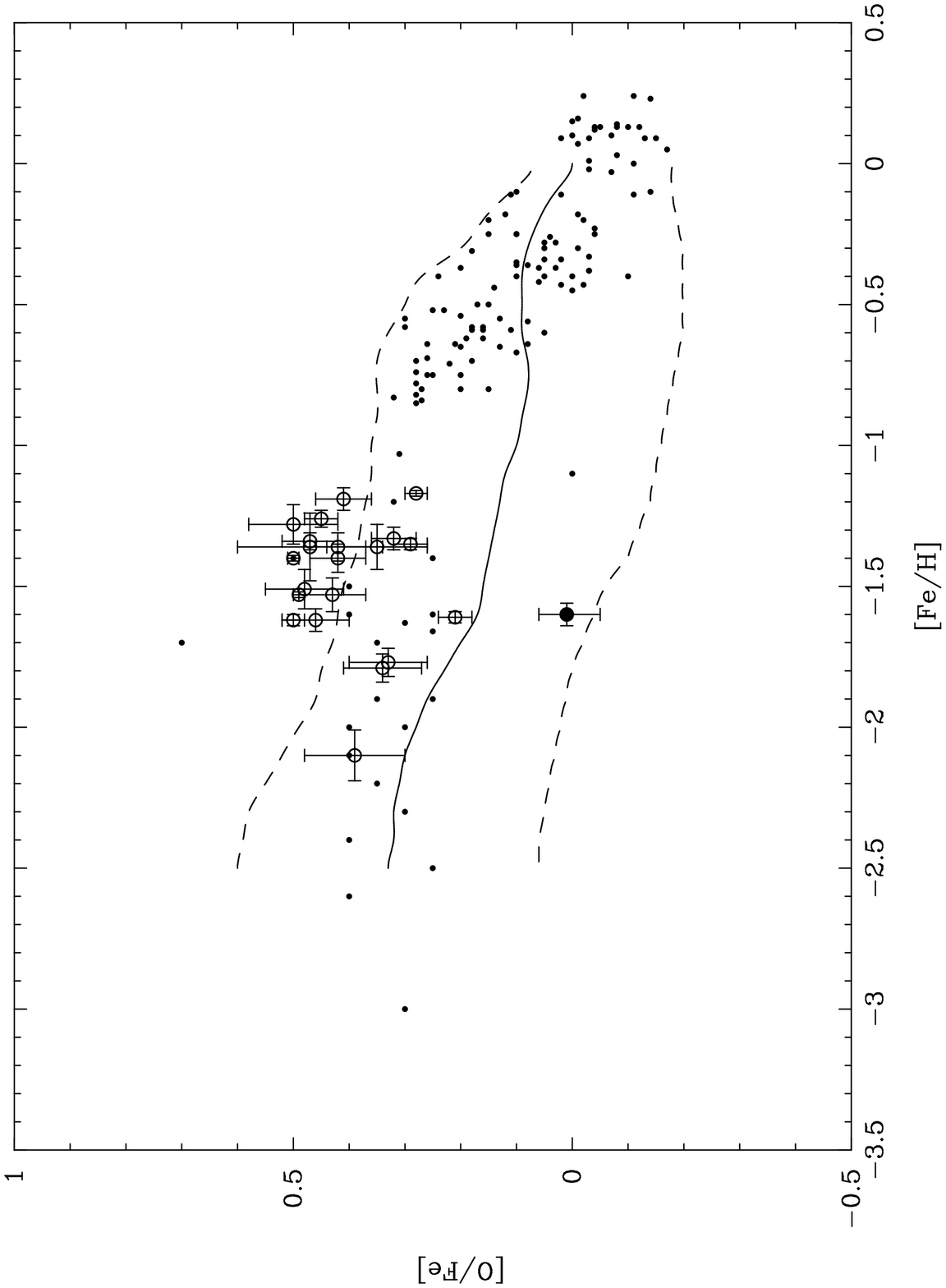]{Location of SBS 0335--052 (filled circle) on the diagram 
of oxygen-to-iron abundance ratio vs. iron abundance. Here [X]$\equiv$
log X -- log X$_\odot$. For comparison, we also shown the data for 
low-metallicity BCGs (Thuan, Izotov \& Lipovetsky 1995, Izotov, Thuan \&
Lipovetsky 1996; open circles), for disk and halo stars (points) from 
Edvardsson et al. (1993), Barbuy (1988) and 
Barbuy \& Erdelyi-Mendes (1989). Solid line is
the [O/Fe] vs. [Fe/H] predicted by the chemical evolution model for the Galaxy,
dashed lines -- chemical evolution model predictions with iron yields two times
larger and lower than that for the solid line (Timmes et al. 1995). While
there is excellent agreement between the BCG data and that for galactic
halo stars, SBS 0335--052 is a significantly deviant point with [O/Fe] which
is close to solar value.}

\figcaption[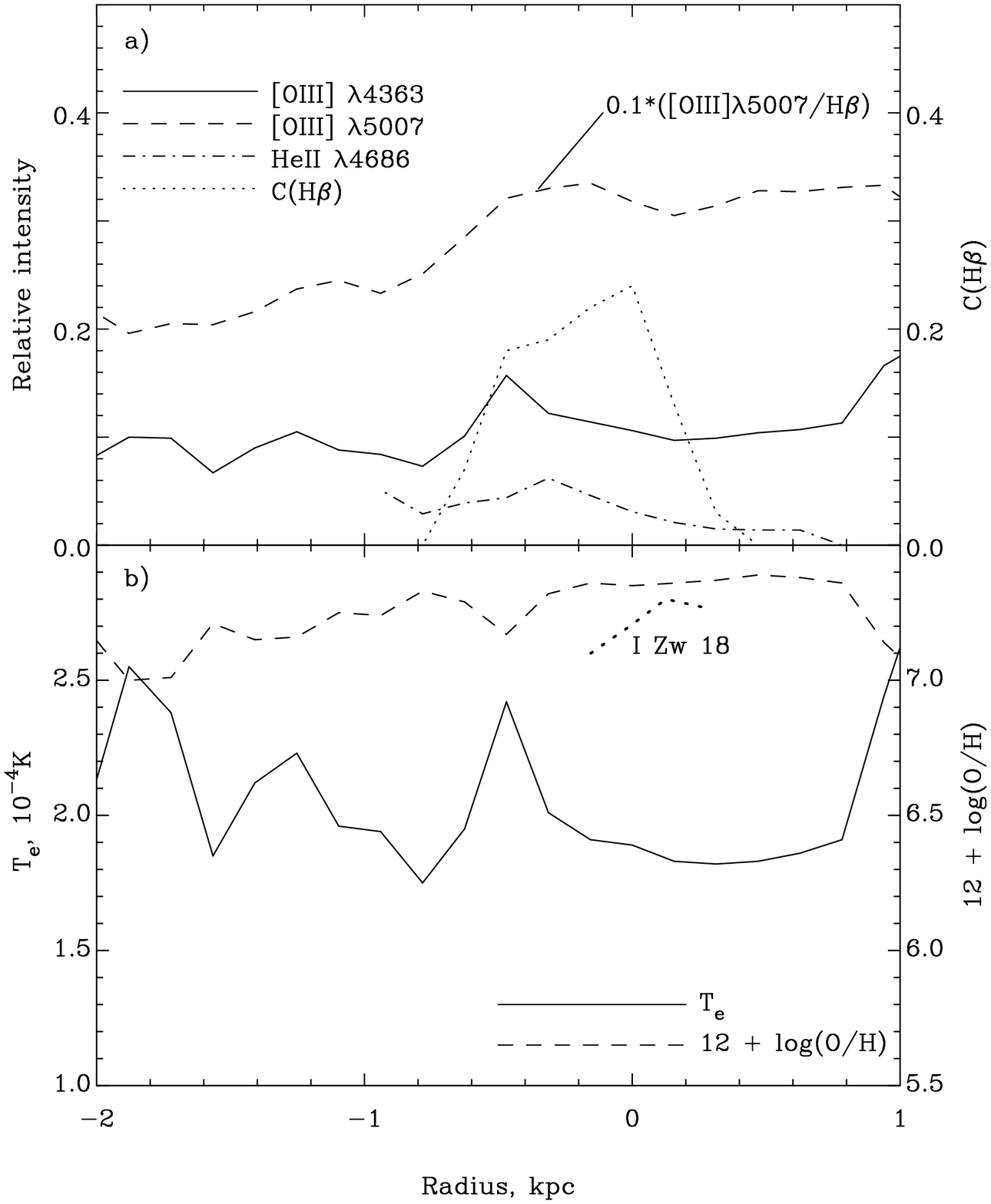]{ a) Distribution along the slit in NW--SE direction 
of [OIII]$\lambda$4363 (solid line), [OIII]$\lambda$5007 (dashed line) 
and HeII $\lambda$4686 (dot-dashed line) 
line intensities as well as extinction coefficient
C(H$\beta$). The intensity of [OIII]$\lambda$5007 emission line is reduced
by a factor of 10. Zero-point is chosen at the maximum in continuum flux 
distribution along the slit. NW direction is to the left, 1$''$ = 261 pc.
Note that [OIII]$\lambda$4363 emission line is observed in region with 
diameter greater 3 kpc, and allows us to measure the electron temperature along
the slit. The maximum in HeII $\lambda$4686 relative intensity distribution is
shifted to NW. We note significant extinction in the brightest part of the
galaxy, which coincides with dust lanes detected by Thuan, Izotov \&
Lipovetsky (1996) on V and I HST images. b) Electron temperature, $T_e$, 
(solid line) and oxygen abundance, 12 + log(O/H), (dashed line) distributions
in the HII region of SBS 0335--052. For comparison, the oxygen abundance 
distribution
in I Zw 18 (Martin 1996) is shown by a dotted line. Note that both electron
temperature and oxygen abundance in SBS 0335--052 are nearly constant across
more than 3 kpc suggesting effective mixing inside the HII region during short
time scale $\leq$10$^7$yr. While the oxygen abundance in SBS 0335--052 
and I Zw 1 is the same, the spatial extent of HII region in SBS 0335--052 is 
several times greater due to a more powerful burst of star formation.}

\figcaption[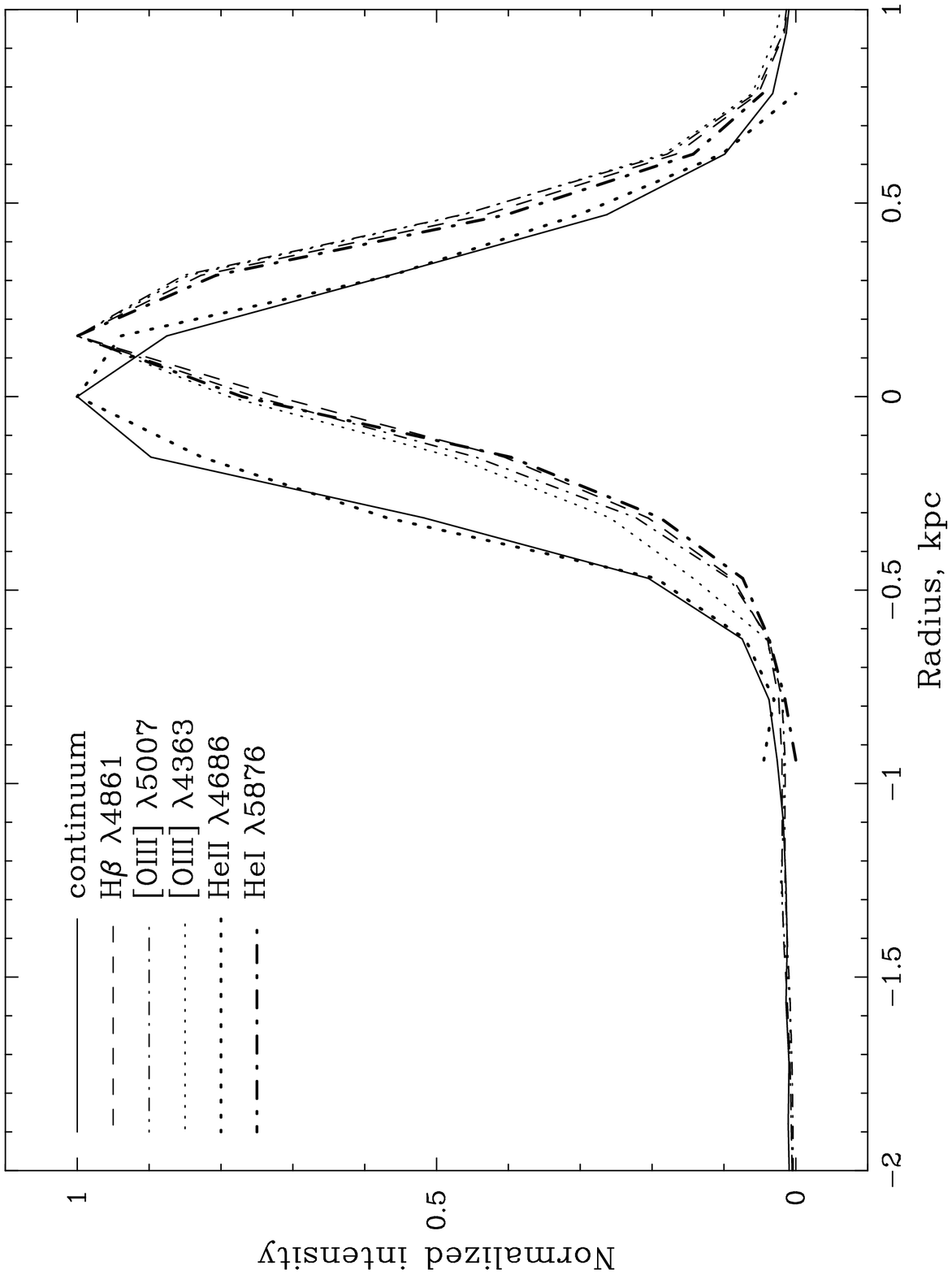]{Spatial distribution along NW--SE direction of continuum and emission
line intensities, normalized to the unity. While the continuum and HeII 
$\lambda$4686 profiles coincide, profiles of other emission lines are shifted
in SE direction and coincide with the youngest stellar cluster detected by
Melnick, Heydari-Malayeri \& Leisy (1992) and Thuan, Izotov \& Lipovetsky 
(1996). Difference in spatial distribution of HeII $\lambda$4686 and other
emission lines implies that ionization of He$^+$ by hard photons with
$\lambda$$\leq$228\AA\ is not caused by main-sequence O-stars.}

\figcaption[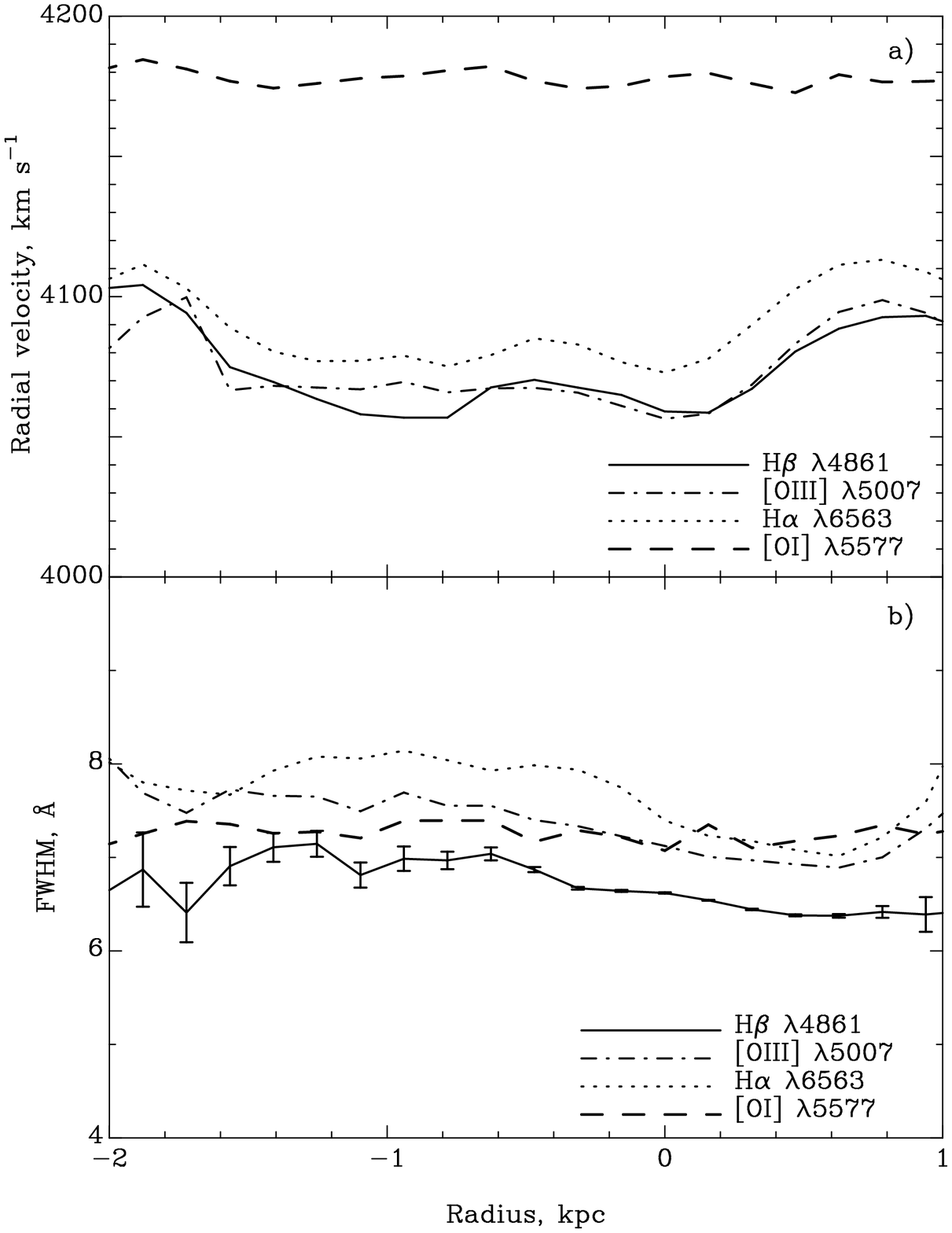]{ a). Spatial radial velocity distribution for strongest emission lines
in SBS 0335--052 in NW--SE direction. For comparison, the spatial wavelength
distribution for night sky emission line [OI]$\lambda$5577 is shown; this line 
does not show significant variations of this line along the slit. We do not
detect a significant gradient in velocity distribution except for a slight
increase of velocity by 50 km s$^{-1}$ in outer parts of HII region. b). 
Distribution of full width at half maximum (FWHM) for strongest emission lines
in SBS 0335--052. For the H$\beta$ $\lambda$4861 emission line width, 
the error bars
are also shown. The FWHM for night sky [OI]$\lambda$5577 emission line is 
shown for comparison. Note that the H$\beta$ $\lambda$4861 line is narrower
than other lines and could be subjected to underlying stellar absorption from
A stars. The gradient in H$\beta$ $\lambda$4861 and [OIII]$\lambda$5007 
emission line width could be explained by turbulent velocity increasing in
NW direction.}

\figcaption[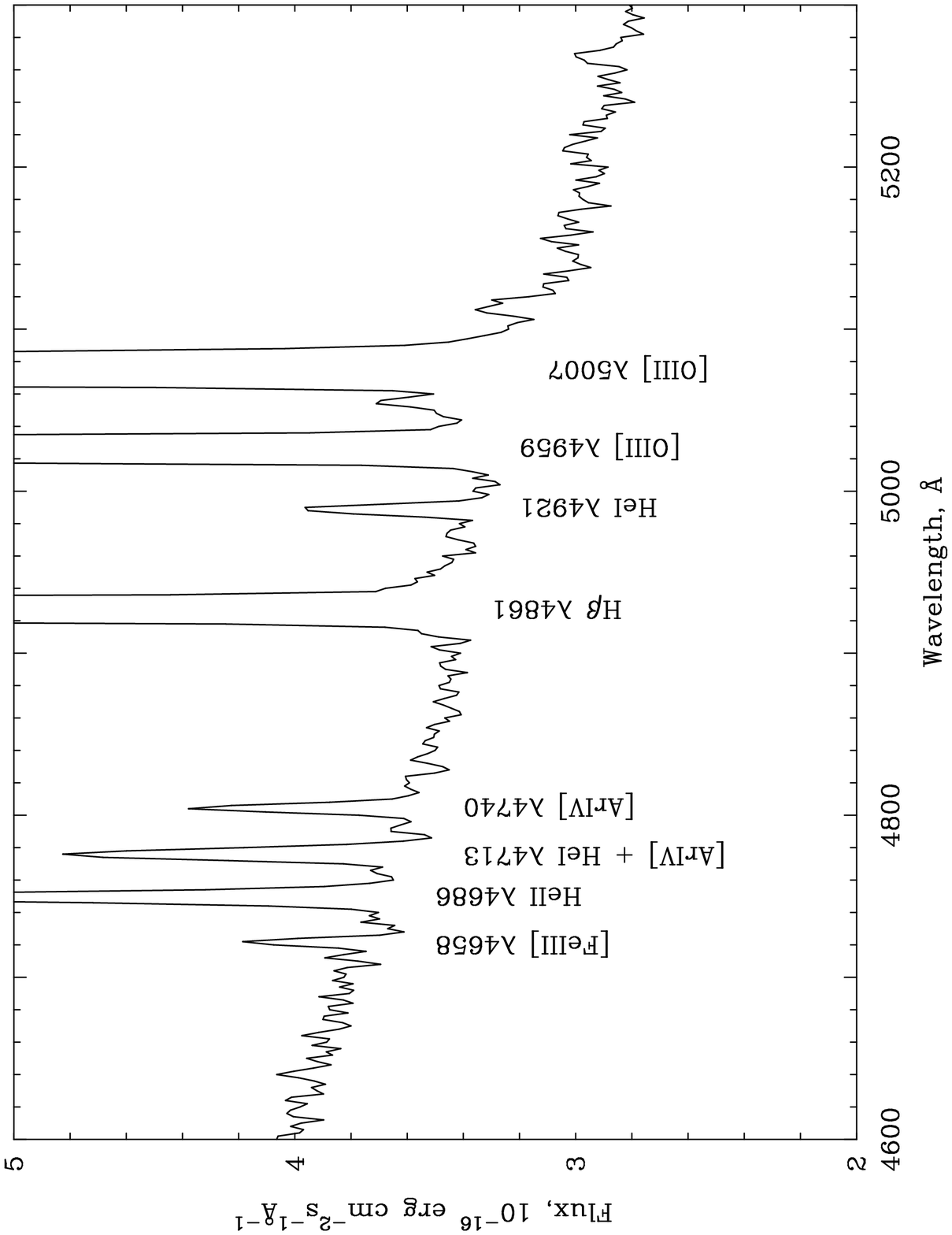]{ Fragment of the SBS 0335--052 spectrum showing the presence of
low-intensity broad component of H$\beta$ $\lambda$4861 emission line with
FWZI $\sim$ 50\AA. Note that HeII $\lambda$4686 emission line is narrow 
implying its nebular origin.}



\clearpage
\begin{table*}
{\small\qub
\begin{tabular}{lccclcc}
  Ion&F($\lambda$)/F(H$\beta$)&I($\lambda$)/I(H$\beta$)&&  Ion&F($\lambda$)/F(H$\beta$)&I($\lambda$)/I(H$\beta$) \\ \tableline
 3727\ [O II]\ ................        &0.195$\pm$0.003&0.233$\pm$0.004&&  4713\ [Ar IV] + He I\ ..              &0.017$\pm$0.002&0.017$\pm$0.002  \\
 3750\ H12\ ...................        &0.021$\pm$0.002&0.038$\pm$0.005&&  4740\ [Ar IV]\ ..............         &0.009$\pm$0.001&0.009$\pm$0.001 \\
 3771\ H11\ ...................        &0.026$\pm$0.002&0.044$\pm$0.005&&  4861\ H$\beta$\ ....................  &1.000$\pm$0.006&1.000$\pm$0.006 \\
 3798\ H10\ ...................        &0.041$\pm$0.002&0.062$\pm$0.005&&  4959\ [O III]\ ...............        &1.076$\pm$0.006&1.054$\pm$0.006 \\
 3835\ H9\ .....................       &0.056$\pm$0.003&0.079$\pm$0.005&&  5007\ [O III]\ ...............        &3.245$\pm$0.015&3.155$\pm$0.014 \\
 3868\ [Ne III]\ ..............        &0.205$\pm$0.003&0.239$\pm$0.004&&  5876\ He I\ ...................       &0.115$\pm$0.002&0.100$\pm$0.002 \\
 3889\ He I + H8\ .........            &0.136$\pm$0.003&0.172$\pm$0.004&&  6300\ [O I]\ ..................       &0.008$\pm$0.001&0.007$\pm$0.001 \\
 3968\ [Ne III]+H7\ .......            &0.189$\pm$0.003&0.230$\pm$0.004&&  6312\ [S III]\ ................       &0.007$\pm$0.001&0.006$\pm$0.001 \\
 4026\ He I\ ...................       &0.012$\pm$0.002&0.013$\pm$0.002&&  6563\ H$\alpha$\ .................... &3.383$\pm$0.015&2.745$\pm$0.013 \\
 4101\ H$\delta$\ .....................&0.217$\pm$0.003&0.255$\pm$0.004&&  6584\ [N II]\ .................       &0.009$\pm$0.002&0.007$\pm$0.001 \\
 4340\ H$\gamma$\ .................... &0.433$\pm$0.003&0.476$\pm$0.004&&  6678\ He I\ ...................       &0.033$\pm$0.001&0.027$\pm$0.001 \\
 4363\ [O III]\ ...............        &0.102$\pm$0.002&0.109$\pm$0.002&&  6717\ [S II]\ ..................      &0.024$\pm$0.001&0.019$\pm$0.001 \\
 4471\ He I\ ...................       &0.033$\pm$0.002&0.035$\pm$0.002&&  6731\ [S II]\ ..................      &0.021$\pm$0.002&0.017$\pm$0.001 \\
 4658\ [Fe III]\ ..............        &0.003$\pm$0.001&0.003$\pm$0.001&&  7065\ He I\ ...................       &0.051$\pm$0.001&0.039$\pm$0.001 \\
 4686\ He II\ .................        &0.028$\pm$0.002&0.028$\pm$0.002&&  7136\ [Ar III]\ ..............        &0.019$\pm$0.001&0.014$\pm$0.001 \\ \\
 C(H$\beta$) dex\ .................&\multicolumn {5}{c}{ 0.27} \\
 F(H$\beta$)\tablenotemark{a}\ ........................ &\multicolumn {5}{c}{  6.06} \\
 EW(H$\beta$)\ \AA\ ................ &\multicolumn {5}{c}{ 178} \\
 EW(abs)\ \AA\ ................ &\multicolumn {5}{c}{ 1.4} \\ 
\tablenotetext{a}{in units of 10$^{-14}$ ergs\ s$^{-1}$cm$^{-2}$}
\end{tabular}
}
\caption{Emission line intensities.}
\end{table*}

\clearpage
\begin{table*}

\begin{center}
{\small\qub
\begin{tabular}{lc} 
Property&Value \\ \tableline
 $T_e$(OIII) (K)\ .............              &19,200$\pm$ 200 \\
 $T_e$(OII) (K)\ \,..............            &15,400$\pm$ 200 \\
 $T_e$(SIII) (K)\ ..............             &17,700$\pm$ 200 \\
 $N_e$(SII) (cm$^{-3}$)\ .........           &  ~~390$\pm$ 10 \\ \\
 O$^+$/H$^+$ ($\times$10$^5$)\ .........     &  0.19$\pm$0.01 \\
 O$^{++}$/H$^+$ ($\times$10$^5$)\ .......    &  1.96$\pm$0.05 \\
 O/H ($\times$10$^5$)\ ..............        &  2.16$\pm$0.05 \\
 12+log(O/H)\ ............                   &  7.33$\pm$0.01 \\ 
 N$^+$/H$^+$ ($\times$10$^7$)\ .........     &  0.49$\pm$0.02 \\
 ICF(N)\ \,.....................             &11.17$\pm$0.08~~ \\
 log(N/O)\ \,..................              &$ -1.59$$\pm$0.03~~ \\ 
 Ne$^{++}$/H$^+$ ($\times$10$^6$)\ .....     &  3.04$\pm$0.10 \\
 ICF(Ne)\ \,....................             & 1.10$\pm$0.01 \\
 log(Ne/O)\ .................                &$ -0.81$$\pm$0.02~~ \\ 
 S$^+$/H$^+$ ($\times$10$^7$)\ \,.........   &  0.37$\pm$0.02 \\
 S$^{++}$/H$^+$ ($\times$10$^7$)\ \,.......  &  1.92$\pm$0.28 \\
 ICF(S)\ \,......................            & 2.65$\pm$0.02 \\
 log(S/O)\ ...................               &$ -1.56$$\pm$0.03~~ \\ 
 Ar$^{++}$/H$^+$ ($\times$10$^7$)\ .....     &  0.40$\pm$0.02 \\
 Ar$^{+++}$/H$^+$ ($\times$10$^7$)\ ...      &  0.77$\pm$0.11 \\
 ICF(Ar)\ \,....................             & 1.01$\pm$0.01 \\
 log(Ar/O)\ .................                &$ -2.26$$\pm$0.04~~ \\ 
 Fe$^{++}$/H$^+$ ($\times$10$^7$)\ \,.....   &  0.58$\pm$0.20 \\
 ICF(Fe)\ \,....................             &13.96$\pm$0.10~~ \\
 log(Fe/O)\ \,.................              &$ -1.43$$\pm$0.04~~ \\ 
\end{tabular}
}
\end{center}
\caption{Ionic and total heavy element abundances.}
\end{table*}

\clearpage
\begin{table*}

\begin{center}
\begin{tabular}{lc} 
Property&Value \\ \tableline
 $N_e$(HeII) (cm$^{-3}$)\ ...           & 152 \\
 $\tau$(HeI $\lambda$3889)\ ........    & 1.5 \\
 (1+$\gamma$)(3889)\ ..........         & 0.956$\pm$0.002 \\
 (1+$\gamma$)(4471)\ ..........         & 1.052$\pm$0.001 \\
 (1+$\gamma$)(5876)\ ..........         & 1.082$\pm$0.002 \\
 (1+$\gamma$)(6678)\ ..........         & 1.022$\pm$0.001 \\
 (1+$\gamma$)(7065)\ ..........         & 1.831$\pm$0.005 \\
 y$^+$(4471)\ ...............           & 0.072$\pm$0.004 \\
 y$^+$(5876)\ ...............           & 0.079$\pm$0.001 \\
 y$^+$(6678)\ ...............           & 0.079$\pm$0.003 \\
 y$^+$(mean)\ ..............            & 0.078$\pm$0.001 \\
 y$^{++}$\ .......................      & 0.003$\pm$0.001 \\
 $\eta$$^d$\ .......................... & 0.539 \\
 ICF(He)\ \,................            & 1.003$\pm$0.020 \\
 y\ \,...........................       & 0.081$\pm$0.002 \\
 Y\ ...........................         & 0.245$\pm$0.006 \\
\end{tabular}
\end{center}

\caption{Helium abundance.}

\end{table*}

\clearpage
\begin{table*}

\begin{center}
\begin{tabular}{lccccccc}
Color&Observations&\multicolumn{3}{c}{Pure gaseous emission}&&\multicolumn{2}
{c}{Instantaneous burst\tablenotemark{a}} \\ \cline{3-5} \cline{7-8}
    & &\multicolumn{1}{c}{I\tablenotemark{b}}&\multicolumn{1}{c}{II\tablenotemark{c}}&\multicolumn{1}{c}{III\tablenotemark{d}}&
&\multicolumn{1}{c}{log t = 6.5}&\multicolumn{1}{c}{log t = 8.0} \\ \tableline
U--B\ ....................&   ...        &--1.3 &--1.1 &--0.9 &&--1.3&0.0 \\
B--V\ ....................&   ...        &  ~0.4&  ~0.5&  ~0.5&&--0.1&0.4 \\
R--I\ ......................&--0.5$\div$0.0\tablenotemark{e}~
&  ~0.2&--0.1&--0.4&&  ~0.0&0.3 \\
V--I\ ......................&  0.0$\div$0.2\tablenotemark{f}&  ~0.5&  ~0.2&--0.1&&--0.1&0.7 \\ \\
\end{tabular}
\end{center}
\tablenotetext{a}{Leitherer \& Heckman 1995.}
\tablenotetext{b}{Colors for gaseous continuum at $T_e$=20000K.}
\tablenotetext{c}{Colors for gaseous emission with EW(H$\beta$)=170\AA\ and
I([OIII]$\lambda$4959)/I(H$\beta$) = 0.6.}
\tablenotetext{d}{Colors for gaseous emission with EW(H$\beta$)=350\AA\ and
I([OIII]$\lambda$4959)/I(H$\beta$) = 0.6.}
\tablenotetext{e}{Lipovetsky et al. 1996.}
\tablenotetext{f}{Thuan, Izotov \& Lipovetsky 1996.}

\caption{Observed and theoretical colors for extended underlying component.}

\end{table*}



\clearpage


\begin{references}

\reference{}  Aller, L.H. 1984, Physics of Thermal Gaseous Nebulae (Dordrecht: Reidel)
\reference{}  Barbuy, B. 1988, \aap, 191, 121
\reference{}  Barbuy, B., \& Erdelyi-Mendes, M. 1989, \aap, 214, 239
\reference{}  Bergeron, J. 1977, \apj, 211, 62
\reference{}  Bessel, M.S. 1990, \pasp, 102, 1181
\reference{}  Bi\'emont, E., Baudoux, M., Kurucz, R.L., Ansbacher, W., \&
Pinnington, E.H. 1991, A\&A, 249, 539
\reference{}  Brocklehurst, M. 1971, \mnras, 153, 471
\reference{}  ---------.\ 1972, \mnras, 157, 211
\reference{}  Campbell, A. 1988, \apj, 335, 644
\reference{} Campbell, A., Terlevich, R., \& Melnick, J. 1986, \mnras, 223, 811
\reference{}  Carigi, L., Colin, P., Peimbert, M., \& Sarmiento, A. 1995, \apj, 
445, 98
\reference{} Castelli, E. \& Kurucz, R.L. 1994, \aap, 281, 817
\reference{} Charlot, S., \& Fall, S.M. 1993, \apj, 415, 580
\reference{} Edvardsson, B., Andersen, J., Gustafsson, B., Lambert, D.L., Nissen, 
P.E., \& Tomkin, J. 1993, \aap, 275, 101
\reference{} Garnett, D. 1989, \apj, 345, 282
\reference{} ---------.\ 1992, \aj, 103, 1330
\reference{}  Garnett, D.R., Kennicutt, R.C., Jr., Chu, Y.-H., Skillman, E.D.
1991, \apj, 373, 458
\reference{}  Garnett, D.R., Skillman, E.D., Dufour, R.J., Peimbert, M., 
Torres-Peimbert, S., Terlevich, R., Terlevich, E., \& Shields, G.A. 1995, \apj,
443, 64
\reference{} Hannaford, P., Lowe, R.M., Grevesse, N., \& Noels, A. 1992, \aap,
259, 301
\reference{} Holweger, H., Bard, A., Kock, A., \& Kock, M. 1991, \aap, 249, 545
\reference{} Izotov, Y.I., Dyak, A.B., Chaffee, F.H., Foltz, C.B., Kniazev, A.Y.,
\& Lipovetsky, V.A. 1996, \apj, 458, 524
\reference{}  Izotov, Yu.I., Lipovetsky, V.A., Guseva, N.G., \& 
Stepanian, J.A. 1990a, Soviet Astr.Lett., 16, 258
\reference{}  Izotov, Yu.I., Lipovetsky, V.A., Guseva, N.G., Kniazev, A.Yu., \& 
Stepanian, J.A. 1990b, \nat, 343, 238
\reference{}  Izotov, Yu.I., Thuan, T.X., \& Lipovetsky, V.A. 1994, \apj, 435, 647
\reference{} ---------. 1996, \apjs, in press
\reference{}  Kingdon, J. \& Ferland, G.J. 1995, \apj, 442, 714
\reference{}  Kunth, D., Maurogordato, S., \& Vigroux, L. 1988, \aap, 204, 10
\reference{}  Kunth, D., \& Sargent, W.L.W. 1983, \apj, 273, 81
\reference{}  ---------.\ 1986, \apj, 300, 496
\reference{} Leitherer, C., \& Heckman, T.M. 1995, \apjs, 96, 9
\reference{} Lequeux, J., Peimbert, M., Rayo, J.F., Serrano, A., \& 
Torres-Peimbert, S. 1979, \aap, 80, 155
\reference{} Lipovetsky, V.A., Chaffee, F.H., Foltz, C.B., Izotov, Yu.I., 
Kniazev, A.Yu., \& Hopp, U. 1996, \apj, in press
\reference{} Loose, H.-H., \& Thuan, T.X. 1985, in Star-Forming Dwarf 
galaxies and related objects, ed. D. Kunth, T.X.Thuan \& J.T.T.Van 
(Gif-sur-Yvette:  Editions Frontieres), 73
\reference{}  Martin, C. 1996, \apj, 465, 680
\reference{}  Melnick, J., Heydary-Malayeri, M., \& Leisy, P. 1992, \aap, 253, 16
\reference{} Pakull, M.W., \& Angebault, L.P. 1986, \nat, 322, 511
\reference{} Pettini, M., Lipman, K., Hunstead, R.W. 1995, \apj, 451, 100
\reference{} Pustilnik, S.A., Lipovetsky, V.A., Izotov, Yu.I., Brinks, E.,
Thuan, T.X., Kniazev, A. Yu., Neizvestny, S.I., \& Ugryumov, A.V. 1996, 
Soviet AJ, submitted
\reference{} Robbins, R.R. 1968, \apj, 151, 511
\reference{} Roy, J.-R., Aube, M., McCall, M.L., \& Dufour, R.J. 1992, \apj, 386, 498
\reference{}  Roy, J.-R., \& Kunth, D. 1995, \aap, 294, 432
\reference{}  Schaerer, D. 1996, preprint
\reference{}  Schaerer, D., \& de Koter, A. 1996, \aap, in press
\reference{} Searle, L., Sargent, W.L.W., \& Bagnuolo, W.G. 1973, \apj, 179, 427
\reference{} Skillman, E., \& Kennicutt, R.C., Jr. 1993, \apj, 411, 655
\reference{} Smits, D.P. 1996, \mnras, 278, 683
\reference{} Stasinska, G. 1990, \aaps, 83, 501
\reference{} Terlevich, E., Terlevich, R., Skillman, E., Stepanian, J., \&
Lipovetsky, V. 1992, in Elements and the Cosmos, eds. M.G.Edmunds \& R. 
Terlevich (Cambridge Univ. Press) 21
\reference{} Terlevich, R., Melnick, J., Masegosa, J., Moles, M., \& Copetti,
M.V.F. 1991, \aaps, 91, 285
\reference{} Thuan, T.X., Izotov, Yu.I., \& Lipovetsky, V.A. 1995, \apj, 445, 108
\reference{}  ---------.\ 1996, \apj, in press
\reference{} Thuan, T.X., Brinks, E., Pustilnik, S.A., Lipovetsky, V.A., \& 
Izotov, Yu.I. 1996, \apj, in preparation
\reference{} Timmes, F.X., Woosley, S.E., \& Weaver, T.A. 1995, \apjs, 98, 671
\reference{} Vigroux, L., Stasinska, G., \& Comte, G. 1987, \aap, 172, 15
\reference{} Weaver, T.A., \& Woosley, S.E. 1993, Phys. Rep., 227, 65
\reference{}  Whitford, A.E. 1958, \aj, 63, 201
\reference{} Woosley, S.E., \& Weaver, T.A. 1995, \apjs, 101, 181
\end{references}
\end{document}